\documentclass[12pt,subeqn,a4paper]{article}

\usepackage{amssymb,amsmath,amsfonts,amsthm, amscd, mathrsfs,helvet}
\usepackage{bm}
\usepackage{graphicx,verbatim}
\usepackage[all]{xy}
\usepackage{xcolor}
\usepackage{braket}
\usepackage{enumitem}
\usepackage[hidelinks]{hyperref}

\usepackage{appendix}

\def\r2{\sqrt{2}}

\newcommand{\B}{\mathcal{B}}
\newcommand{\nn}{\nonumber}

\newcommand{\ent}{\mathcal{S}}

\newcommand{\ncAdS}{X}

\newcommand{\Sp}{C}

\newcommand{\Hosc}{\Delta}

\newcommand{\Q}{\mathbb{Q}}
\renewcommand{\S}{\mathbb{S}}

\begin{document}
\newcommand{\nd}[1]{/\hspace{-0.5em} #1}
\begin{titlepage}
\begin{flushright}
{\bf July 2022} \\ 
\end{flushright}
\begin{centering}
\vspace{.2in}
 {\large {\bf Black Hole Entropy from Quantum Mechanics}}

\vspace{.3in}

Nick Dorey, Rishi Mouland and Boan Zhao\\
\vspace{.1 in}
DAMTP, Centre for Mathematical Sciences \\ 
University of Cambridge, Wilberforce Road \\ 
Cambridge CB3 0WA, UK \\
{\tt N.Dorey@damtp.cam.ac.uk, r.mouland@damtp.cam.ac.uk, bz258@cam.ac.uk} \\
\vspace{.2in}
%
%
\vspace{.4in}
{\bf Abstract} \\

We provide evidence for a holographic duality between superconformal quantum mechanics on the moduli space of Yang-Mills instantons and M-theory in certain asymptotically 
$AdS_{7}\times S^{4}$ backgrounds with a plane-wave boundary metric. We show that the gravitational background admits a supersymmetric black hole solution whose entropy is precisely reproduced by the superconformal index of the dual quantum mechanics.   
\end{centering}

\end{titlepage}

\setcounter{tocdepth}{2}


\vspace{2em}
\section*{Introduction}

A key task for any candidate theory of quantum gravity is to provide a microscopic explanation for the thermodynamic properties of black holes.  String theory provides a concrete setting in which black hole microstates can be counted, reproducing the classical Bekenstein-Hawking entropy formula, at least for supersymmetric black holes \cite{StromingerVafa}. The AdS/CFT correspondence yields further insight by providing a dual description of black hole microstates in quantum field theory. In particular, recent work has lead to rapid progress in accounting for the entropy of supersymmetric black holes in Anti-de-Sitter space (see \cite{Zaf} and references therein). However, all currently understood examples where this analysis is possible involve dual field theories in spacetime dimension $D\geq 2$. Holographic dualities involving one-dimensional theories or, in other words, finite-dimensional quantum mechanics, are still poorly understood \cite{frag} and there are few working examples. A notable exception is the BFSS matrix model \cite{BFSS} which is holographically dual to the near horizon geometry of D0 branes in Type IIA string theory. Such finite dimensional examples are particularly interesting conceptually as they have a fully non-perturbative definition. This is also important in practical terms as as they can be simulated on a computer\footnote{For a survey of recent progress and prospects in simulations of holographic matrix quantum mechanics on classical and quantum computers see \cite{numerical}.} avoiding many of the difficulties associated with lattice field theory in higher dimensions.
\paragraph{}
So far there has been little intersection between the two strands of research described above. In this paper we propose a precise duality between a superconformal quantum mechanical model and a gravitational theory which admits a family of supersymmetric black hole solutions.  
We test the duality by computing the index degeneracies of BPS states from first principles in quantum mechanics and find an exact match to the Bekenstein-Hawking entropy of the black hole. As far as we know, this is the first example where black hole entropy can be reproduced analytically in a finite-dimensional model. The model therefore potentially provides a novel setting where analytical and numerical approaches may coexist.
\paragraph{}
The correspondence described here between quantum mechanics and gravity is self-contained and can be stated without reference to field theory or string/M-theory. However it is also related to the more familiar holographic duality between the six dimensional $(2,0)$ CFT  and M-theory on $AdS_{7}\times S^{4}$. In particular, building on the ideas of \cite{Maldacena:2008wh}, it corresponds to a special limit in which the boundary theory 
reduces to quantum mechanics. This will be explored further in two follow up papers \cite{GravityFollowUp,IndexFollowUp}. In the rest of the paper we provide an overview of our results. Additional details are provided in a series of appendices.


\section*{The quantum mechanical model}

We consider superconformal quantum mechanics \cite{Fubini,Eliezer} on the ADHM moduli space, $\mathcal{M}_{K,N}$, of $K$ Yang-Mills instantons in a $U(N)$ gauge theory defined on $\mathbb{R}^{4}$. The moduli space is a hyperK\"{a}hler cone of complex dimension $d_{\mathbb{C}}=2KN$. 
A $(0+1)$-dimensional supersymmetric non-linear $\sigma$-model with target space $\mathcal{M}_{K,N}$ therefore admits an 
$\mathfrak{osp}(4^{*}|4)$ superconformal symmetry \cite{Singleton1}. The theory we seek has a discrete\footnote{As usual in superconformal quantum mechanics we are working in a Hilbert space appropriate for diagonalising the dilatation operator $D$ rather than the Laplacian denoted $H$. By a standard argument this is equivalent to the spectrum of the operator $\Hosc=H+\Sp$ where $\Sp$ is certain a harmonic potential on $\mathcal{M}_{K,N}$, yielding a discrete spectrum. See Appendix A.1 for further details.} spectrum consisting of unitary, irreducible, lowest-weight representations of this algebra analogous to the spectrum of local operators described by  radial quantization of a higher dimensional CFT.   
As $\mathcal{M}_{K,N}$ is a singular space, a proper definition of the theory requires further elaboration. The instanton moduli space has a smooth resolution $\tilde{\mathcal{M}}_{K,N}$ characterised by a real parameter  
$\xi_{\mathbb{R}}$ with the dimensions of length. Although the resolution breaks the superconformal symmetry described above, it preserves ordinary supersymmetry as well as (most of) the global and $R$-symmetries of the model. As discussed in \cite{SingletonDorey, SingletonThesis,DoreyBarnsGraham, BarnsGrahamThesis}, supersymmetric quantum mechanics on the smooth manifold $\tilde{\mathcal{M}}_{K,N}$ equipped with a certain potential provides a regulated version of the model. We expect to recover an  $\mathfrak{osp}(4^{*}|4)$ invariant spectrum in the ``continuum'' limit $\xi_{\mathbb R} \rightarrow 0$. We also note that the both the superconformal model and its regulated counterpart can be embedded in a manifestly finite matrix quantum mechanical gauge theory. 
The latter model can be put in a form suitable for numerical simulation on a computer\footnote{Indeed computer simulations of closely related models have already appeared \cite{Simulation}.}.            
\paragraph{} 
States in the model are labelled by a set of Cartan generators for $\mathfrak{osp}(4^{*}|4)$ denoted $\Delta$, $J_{1}+J_{2}$, $Q_{1}$, $Q_{2}$. Here the quantum number\footnote{We will use the same notation throughout to denote both a Cartan generator and its eigenvalue.} $\Delta$ is a scaling dimension which obeys a BPS bound of the form, 
\begin{eqnarray}
\Delta & \geq & J_{1}+J_{2}+2Q_{1}+2Q_{2} 
\label{bps1}
\end{eqnarray}
Although the full spectrum of the model is out of reach, analytic methods can provide some control over the spectrum of BPS states which saturate the bound. As in more familiar higher-dimensional examples \cite{Romel,Kinney}, these states belong to special short multiplets of the superconformal algebra and they can be counted by defining a {\em superconformal index} \cite{SingletonDorey} (see also \cite{KimLee}). It is useful to define the following linear combinations of the charges, 
\begin{eqnarray}
L_{t} & := & J_{1}+J_{2}+Q_{1}+Q_{2} \nonumber \\
L_{y} & := & Q_{1}-Q_{2} \nonumber 
\nonumber
\end{eqnarray}
which commute with the supercharges used to define the index. The remaining linear combination $F:=-2Q_{2}$ instead plays the role of fermion number. 
In addition to superconformal invariance, the theory also has an $\mathfrak{su}(2)\oplus\mathfrak{su}(N)$ global symmetry with Cartan generators denoted $L_{x}:=J_{1}-J_{2}$ and $n_{a}$, $a=1,2\ldots,N-1$. 
The resulting index corresponds to the following trace over the space of BPS states, 
\begin{eqnarray}
\mathcal{I}_{\rm SC}[\mathcal{M}_{K,N}] &  := & {\rm Tr}_{\rm BPS} \left[(-1)^{F} t^{L_{t}}x^{L_{x}}y^{L_{y}}\prod_{a=1}^{N}w_{a}^{n_{a}}\right] 
\label{trace}
\end{eqnarray}
\paragraph{}
Importantly the same index can also be defined in the regulated version of the model, where it corresponds to the Witten index of supersymmetric quantum mechanics on the resolved space, with a potential corresponding to the norm of a holomorphic Killing vector \cite{Potentials}. The resulting index is independent of the resolution parameter $\xi_{\mathbb{R}}$ and serves as a regulated definition of the superconformal index. It also has a nice geometric interpretation as an Euler characteristic for equivariant sheaf cohomology on the resolved space \cite{DoreyBarnsGraham}.
\begin{eqnarray}
\mathcal{I}_{\rm SC} \left[\mathcal{M}_{K,N}\right] & = & \sum_{p=0}^{d_{\mathbb{C}}} \left(-\frac{y}{t}\right)^{p-d_{\mathbb{C}}/2}\, \chi\left(\tilde{\mathcal{M}}_{K,N},\mathcal{A}^{p}\right)   
\label{sheaf}
\end{eqnarray}
Here $\chi(Y,\mathcal{A}^{p})$ denotes the equivariant Euler character of the sheaf of $p$-forms on a smooth variety $Y$. The latter can be computed by standard localisation theorems in equivariant K-theory giving a closed formula for the index as a sum over fixed points of the maximal torus of a group action on $\tilde{\mathcal{M}}_{K,N}$.
These are enumerated by $N$-coloured partitions \cite{nakajimabook} $\vec{\lambda}\in \mathcal{P}^{N}$ of total weight $||\vec{\lambda}||=K$. The resulting expression for the superconformal index is,  
\begin{eqnarray}
\mathcal{I}_{\rm SC}\left[\mathcal{M}_{K,N}\right] & = & \sum_
{\{\lambda\in \mathcal{P}^{N}:\,\, ||\vec{\lambda}||=K\}}\,\, \prod_{i,j=1}^{N}\,\, \prod_{s\in Y(\lambda^{(i)})}\,\,{\rm Pexp}\left(\frac{z_{i}}{z_{j}}\,t^{g_{ij}(s)}\,x^{f_{ij}(s)}\left[ty\right] \left[t/y\right]\right) 
\label{loc1}
\end{eqnarray}
where $w_{a}=z_{i+1}/z_{i}$ for $a=i=1,\ldots,N-1$. See Appendix A.2 for definitions and notation. The consistency of this expression with the original definition (\ref{trace}) as  trace over an $\mathfrak{osp}(4^{*}|4)$-invariant spectrum is non trivial. It requires, for example, that the resulting rational function of the fugacities appearing in (\ref{loc1}) has an expansion in positive powers of $t$ with coeficients which are Laurent polynomials in the remaining variables\footnote{This property can be checked to high order by explicit calculation and a proof will be given in \cite{Andypaper}.}        
\paragraph{}
The index computed above coincides exactly with the $K$-instanton contribution to the Nekrasov partition function of an auxiliary five-dimensional $U(N)$ gauge theory, a fact we will use extensively below.  In this context the $\mathfrak{su}(N)$ symmetry of the model is inherited from the global part of the $U(N)$ gauge group in five dimensions. As usual, for comparison with a gravitational dual, we will focus on the $\mathfrak{su}(N)$ invariant sector of the model. 
We project onto the $\mathfrak{su}(N)$ singlets by integrating over the maximal torus of $SU(N)$ with the Haar measure\footnote{In fact, the agreement we find between the  model and a gravitational dual does not seem to depend on this projection. A related issue in the context of holography has recently been discussed in \cite{MaldacenaMilekhin}.}.   
It is also convenient sum over the instanton number $K$, with fugacity $q_{\tau}=\exp(2\pi i \tau)=\exp(-\beta)$ although ultimately we will project back onto sectors of fixed $K$. The final form of our index is, 
\begin{eqnarray}
\mathcal{I}_{\rm QM}\left[t,x,y,q_{\tau}\right] & = & \sum_{K=0}^{\infty} \,q_{\tau}^{K}\, \int \, 
d\mu_{\rm Haar}[{Z}]\, \, \mathcal{I}_{\rm SC}\left[\mathcal{M}_{K,N}\right] 
\label{indqm} 
\end{eqnarray}
\paragraph{}
As mentioned above, the index at each instanton number has an expansion in positive powers of the complex fugacity $t$.  
The coefficients in this expansion encode information about the degeneracies of BPS states in the model. Concretely, if we define, 
\begin{eqnarray}
\mathcal{C}(L_{t},L_{x},L_{y},K) & \in & \mathbb{Z} \nonumber 
\end{eqnarray}
as the coefficient of, 
\begin{equation}
t^{L_{t}}x^{L_{x}}y^{L_{y}}q_{\tau}^{K}
\label{monintro}
\end{equation}
in this expansion, then the definition of the index as a trace over BPS states given above implies, 
\begin{eqnarray}
\mathcal{C}(L_{t},L_{x},L_{y},K) & = & \sum_{F\in \mathbb{Z}}\, \left(-1\right)^{F}\, 
d(L_{t},L_{x},L_{y},K,F) 
\label{cvsdintro}
\end{eqnarray}
where $d(\ldots)\in \mathbb{Z}_{\geq 0}$ denote the corresponding degeneracies. 
\paragraph{}
Brute force evaluation of the coefficients $\mathcal{C}(L_{t},L_{x},L_{y},K)$ on a 
computer\footnote{Some sample data for the most computationally tractable case $N=1$ is  given is Appendix B. Preliminary data for $N>1$ reveals a qualitatively similar behaviour.} reveals that they grow rapidly in magnitude as we increase $L_{t}$, $L_{x}$ $L_{y}$ and $K$ and oscillate with a sign that is not correlated in any obvious way with the values of these charges. Obtaining an analytic understanding of this growth is one the main goals of this paper. 
The numbers $\mathcal{C}$ can also be related to alternating sums of the dimensions of certain subspaces of fixed grade in the equivariant sheaf cohomology of the resolution $\tilde{\mathcal{M}}_{K,N}$ \cite{DoreyBarnsGraham,Andypaper}.   

\section*{Building the duality}

In order to identify a gravitational dual of the superconformal quantum mechanics defined above, we will first exploit its relation to the $(2,0)$ superconformal field theory in six dimensions. The most familiar form of this relation arises from compactification of the the $(2,0)$ theory of type $A_{N-1}$ on a spacelike circle of radius $R$. The resulting  effective theory in five dimensions is maximally supersymmetric $U(N)$ Yang-Mills theory with coupling $g_{5}^{2}\sim R$. Yang-Mills instantons give rise to classical BPS solitons in this gauge theory which correspond to Kaluza-Klein modes on $S^{1}$ in the original six dimensional description. The quantum mechanical $\sigma$-model with target space $\mathcal{M}_{K,N}$ considered above then provides a low-energy effective description of the sector of the $(2,0)$ theory with $K$ units of momentum on $S^{1}$.
\paragraph{}
 A more profound relation between the two theories emerges when we perform a large boost along the compactified direction (together with a suitable rescaling). In the Seiberg limit \cite{Seiberg} where the parameter of the boost becomes infinite, the low-energy description becomes exact. The result is a precise duality between the $(2,0)$ theory compactified on a null circle with $K$ units of momentum, and superconformal quantum mechanics on $\mathcal{M}_{K,N}$. A key point underlying this duality is the fact that (the simple part of) the stabilizer of the six-dimensional $(2,0)$ superconformal algebra under such a null compactification coincides with the quantum mechanical superconformal algebra $\mathfrak{osp}(4^{*}|4)$. In this context, the Cartan generators $J_{1}$ and $J_{2}$ correspond to angular momenta in two orthogonal planes transverse to the lightcone, while $Q_{1}$ and $Q_{2}$ correspond to R-charges. The eigenvalues of the Hamiltonian $\Hosc=H+\Sp$ coincide with scaling dimensions under $D$, which is the the so-called lightcone or Lifschitz dilatation operator of the $(2,0)$ theory.   
Such null compactifications are examples of Discrete Light Cone Quantization (DLCQ), and the fact that they lead to finite dimensional non-relativistic quantum mechanics for fixed  values of the null momentum $K$ is familiar in this context. Indeed the model considered here was first introduced \cite{SeibergEva, ABS} as a DLCQ formulation of the $(2,0)$ superconformal field theory. 
\paragraph{}
The operator $\Hosc=H+C$, which is the natural Hamiltonian for superconformal quantum mechanics, does not act in a simple way in Minkowski space. For the present purposes, the duality can be put in a more useful form by a conformal transformation \cite{Duval:1994qye} which maps $\mathbb{R}^{5,1}$ 
to a pp-wave geometry with metric,  
\begin{align}
  ds^2 = - 2 dx^+ dx^- -  x^i x^i ( dx^-)^2 + dx^i dx^i
  \label{ppintro}
\end{align}
with $i=1,\dots,4$, subject to null compactification via the identification $x^{+}\sim x^{+}+2\pi$. 
In these coordinates, $\Hosc$ simply generates translations of $x^{-}$, while $J_1,J_2$ are a pair of commuting rotations in the $x^i$ directions.
Due to the gravitational potential well in the transverse directions of the metric (\ref{ppintro}), the resulting spectrum is discrete and, thanks to the residual superconformal invariance, consists of multiplets of $\mathfrak{osp}(4^{*}|4)$. The map thus provides a lightcone analogue of the usual radial quantization map \cite{Nishida:2007pj,Goldberger:2008vg}. Following \cite{ABS, Maldacena:2008wh}, we then have, 
\paragraph{}
{\bf Duality I:} Superconformal quantum mechanics on $\mathcal{M}_{K,N}$ is equivalent to 
the six-dimensional $(2,0)$ theory of type $A_{N-1}$ in the null-compactified pp-wave geometry (\ref{ppintro}) with $K$ units of null momentum in the compact direction. 
\paragraph{}
We are able to provide an immediate check of this duality at the level of BPS states. As is well known, the six-dimensional pp-wave geometry (without null compactification) can be obtained as a {\em Penrose limit} of the Lorentzian cylinder $S^{5}\times \mathbb{R}_{t}$. 
If we replace the $S^{5}$ factor with a suitable orbifold $S^{5}/\mathbb{Z}_{n}$ and coordinate the usual Penrose limit with the limit $n\rightarrow \infty$, we instead obtain 
the null-compactified version of the pp-wave (\ref{ppintro}) as a limiting geometry. Fortunately, an appropriate supersymmetric index for the $(2,0)$-theory on the orbifold $S^{5}/\mathbb{Z}_{n}$ has already been computed via reduction to 5d gauge theory on a spatial $\mathbb{CP}^{2}$  and subsequent localisation \cite{Kim:2013nva}. The resulting index is expressed as a convolution of contributions from instantons and perturbation theory on three distinct coordinate patches of $\mathbb{CP}^{2}$ as well as a sum over flux sectors. By applying the combined Penrose and $n\rightarrow \infty$ limits we can obtain an index for the $(2,0)$ theory on a null compactified pp-wave background. We find that the orbifold index undergoes dramatic simplification in this limit and coincides exactly with the index $\mathcal{I}_{\rm QM}$ for superconformal quantum mechanics defined above \cite{IndexFollowUp}. 
\paragraph{}
The main motivation for relating our quantum mechanical model to the $(2,0)$ theory is that the latter has a known gravitational dual. The theory of type $A_{N-1}$ is dual to M-theory on $AdS_{7}\times S^{4}$ with $N$ units of four-form flux through $S^{4}$. When $N>>1$, M-theory reduces to eleven-dimensional supergravity on the same background. As noted above,  the six dimensional pp-wave spacetime is related to flat Minkowski space by a conformal transformation. Correspondingly we can find coordinates on $AdS_{7}$ that manifestly realise pp-wave asymptotics; explicitly, we have
\begin{align}
  ds^2_{\text{AdS}} = \frac{dr^2}{g^2r^2} + r^2 \Big(-2dx^+ dx^- - x^i x^i (dx^-)^2 + dx^i dx^i\Big) - \frac{1}{g^2}(dx^-)^2
  \label{eq: AdS7 pp-wave slicing}
\end{align}
where $g=R^{-1}_\text{AdS}$ is the inverse AdS radius. We define $\ncAdS$ as this spacetime subject to the identification $x^+ \sim x^+ + 2\pi$. It is clear then that $\ncAdS$ realises null-compactified pp-wave asymptotics. Then, applying the standard holographic logic and using Duality I, we have our second proposal \cite{Maldacena:2008wh}, 
\paragraph{}
{\bf Duality II:} Superconformal quantum mechanics on $\mathcal{M}_{K,N}$ is equivalent to 
M-theory on spacetimes asymptotic to $\ncAdS\times S^{4}$ . More precisely we restrict to the sector of this theory with with $K$ units of momentum around the null boundary circle, as well as $N$ units of four-form flux through $S^4$.
\paragraph{}
As usual the basic kinematics underlying the duality ensure that the superconformal algebra $\mathfrak{osp}(4^{*}|4)$ of the boundary theory is linearly realised as the algebra of (super)-isometries in the bulk\footnote{In fact, the full group of super-isometries of the bulk is the maximal subgroup of $\mathfrak{osp}(8^*|4)$ preserved by the null compactification. This is the supersymmetrised Schr\"odinger group in four spatial dimensions, of which $\mathfrak{osp}(4^*|4)$ is a subgroup. This full symmetry group is indeed realised in the superconformal quantum mechanical model we construct, albeit non-linearly, as first noted in \cite{Aharony:1997an}.}. The scaling dimension $\Delta$ corresponds to translations in the bulk time coordinate corresponding to boundary lightcone time $x^{-}$.    
The Cartan generators $J_{1}$ and $J_{2}$ correspond to transverse angular momenta in the AdS pp-wave part of the bulk space time while $Q_{1}$ and $Q_{2}$ are associated with rotations on the internal $S^{4}$. 
Furthermore, the parameters on either side of the duality are fixed by the relation
\begin{align}
  \frac{R_\text{AdS}}{l_p} = \frac{2R_{S^4}}{l_p} = 2(\pi N)^{1/3}
  \label{eq: AdS radius in Planck units}
\end{align}
where $l_p$ denotes the 11-dimensional Planck length. 

\section*{A black hole and its entropy}

 One can then seek M-theory solutions with the desired asymptotics. As first proposed in \cite{Maldacena:2008wh}, beginning with some asymptotically AdS$_7\times S^4$ solution, one can take the AdS dual of the Penrose limit discussed above to arrive at solutions with $\ncAdS\times S^4$ asymptotics. We will call this the boundary Penrose limit to distinguish it from the conventional Penrose limit of the bulk theory (ie the BMN limit \cite{BMN}). In more detail, the $(2,0)$ theory on the Lorentzian cylinder $S^{5}\times \mathbb{R}_{t}$ is dual to M-theory on $AdS_{7}\times S^{4}$ in global coordinates. After taking the boundary Penrose limit, the asymptotic geometry takes the form (\ref{eq: AdS7 pp-wave slicing}); a subsequent identification of a null coordinate results in precisely the spacetime geometry $\ncAdS$ defined above. In particular, this procedure was applied in \cite{Maldacena:2008wh} to (non-supersymmetric) black hole solutions in $AdS_{7}\times S^{4}$ to obtain new black hole solutions with $\ncAdS\times S^4$ asymptotics. For brevity, we will refer to any such black hole supergravity solutions with $\ncAdS\times S^4$ asymptotics as an $\ncAdS$ black hole.
\paragraph{}
Crucial to the duality is the compact null direction in the asymptotic geometry $\ncAdS\times S^4$. 
As a null circle is equivalent to a spacelike circle of vanishing size this seems to preclude a description in terms of classical gravity. In particular, the bulk configuration of pure $\ncAdS\times S^4$  should be dual to the vacuum state in the boundary theory. Because the boundary circle remains null throughout the bulk,  
there is no reliable approximation of this state in terms of classical gravity. However, the situation is somewhat better for $\ncAdS$ black holes. Such solutions should be dual to thermal ensembles of (generically non-BPS) states of large charge in the boundary quantum mechanics. Then, as first noted in \cite{Maldacena:2008wh}, the null circle on the boundary becomes spacelike in the bulk, and for sufficiently large momentum along it, this circle is large enough in some regions of the bulk spacetime to allow a classical gravity description. 
\paragraph{}    
As the index of the quantum mechanical model exhibits a degeneracy of BPS states which grows rapidly with the charges, it's natural to seek an explanation in terms of the entropy of a {\em supersymmetric} black hole in the dual theory; that is, an $\ncAdS$ black hole whose charges satisfy the BPS bound for $\mathfrak{osp}(4^{*}|4)$. Fortunately, although such solutions have not been considered before, they can be constructed from known supersymmetric asymptotically AdS$_7$ Kerr-Newman black hole solutions following the method outlined above.
\paragraph{}
The general non-extremal black hole solution in AdS$_7$ is expected to depend on six parameters, which can be exchanged in favour of conserved charges corresponding to the six Cartan generators of $SO(2,6)\times SO(5)$. These are made up of a mass $M$ and three angular momenta $J_1,J_2,J_3$ corresponding respectively to time translations and $S^5$ rotations on the $\mathbb{R}_t\times S^5$ conformal boundary, and two electric charges $Q_1,Q_2$ corresponding to the $SO(5)$ gauge fields arising after Kaluza-Klein reduction on $S^4$. However, the most general solutions currently known are constrained to have either equal angular momenta $J_1=J_2=J_3$ \cite{Chong:2004dy}, or equal charges $Q_1=Q_2$ \cite{Chow:2007ts}. We will focus exclusively on the latter solutions here. Such solutions then admit a BPS limit, in which two degrees of freedom are lost. Firstly, a certain relation amongst parameters ensures that the BPS bound
\begin{align}
  M \ge J_1 + J_2 + J_3 + 4Q
  \label{eq: Chow BPS bound}
\end{align}
is saturated, while an additional non-linear relation amongst $\{J_1,J_2,J_3,Q\}$ is also required to avoid naked closed time-like curves. The entropy $\mathcal{S}_{\text{BH}}>0$ of such BPS black holes is found to be
\begin{align}
\bigg(\frac{\mathcal{S}_{\text{BH}}}{2\pi}\bigg)^2 & =  \frac{Q^{3}-(N^{3}/6)\left(J_{1}J_{2}+J_{2}J_{3}+J_{1}J_{3}\right)}{Q - 
N^{3}/6}
\label{eq: Chow entropy}
\end{align}
and the non-linear constraint mentioned above can be recast as the statement that $\mathcal{S}_{\text{BH}}$ must also satisfy
\begin{align}
  \left(\frac{\mathcal{S}_{\text{BH}}}{2\pi}\right)^4 - 2A \left(\frac{\mathcal{S}_{\text{BH}}}{2\pi}\right)^2 + B=0
\label{s1}
\end{align}
with
\begin{eqnarray} 
A \,=\, \frac{N^{3}}{3}\left(J_{1}+J_{2}+J_{3}\right)\,\,+\,\,3Q^{2}, & \qquad{} & 
B\,=\, \frac{2}{3}N^{3}J_{1}J_{2}J_{3}\,\,+\,\, Q^{4}
\nonumber 
\label{s2}
\end{eqnarray}
Then, following the method of \cite{Maldacena:2008wh} as outlined above, we obtain from Chow's solutions a new class of black holes precisely realising $\ncAdS\times S^4$ asymptotics. Our solutions retain the five degrees of freedom of the initial solution of \cite{Chow:2007ts}, which can be exchanged for a set of charges $\{\Delta,K,J_1,J_2,Q\}$ corresponding to the isometries of $\ncAdS\times S^4$ as detailed above. The fact that Chow's black hole configurations are indeed solutions of the bulk supergravity ensures that our solutions are too; nonetheless, we have independently checked using \textit{Mathematica} that this is indeed the case. Here, we present the key relevant details of our solutions, while their full details can be found in Appendix \ref{app: BH solution}. The derivation of our solutions in the boundary Penrose limit of those of Chow will appear in a forthcoming paper \cite{GravityFollowUp}.
\paragraph{}
 In order to match with the index of the quantum mechanics, we then specialise to  solutions which saturate the BPS bound (\ref{bps1}) for  $\mathfrak{osp}(4^{*}|4)$.
We also find an additional non-linear constraint on the charges which arises as the smooth boundary Penrose limit of the corresponding constraint on Chow's solution described above. We are then left with a three parameter family of supersymmetric $\ncAdS$ black holes.  
To describe the resulting thermodynamics, it's useful work in terms of $K$ and the two linear combinations $L_{t}=J_{1}+J_{2}+2Q$ and $L_{x}=J_{1}-J_{2}$ introduced above. Then, the constraint allows us to express the remaining independent charge as $Q=Q(L_{t},L_{x},K)$, while we can also write the Bekenstein-Hawking entropy as $\ent_\text{BH}=\ent_\text{BH}(L_t,L_x,K)$. These take the forms \begin{align}
  \ent_\text{BH}(L_t,L_x,K) = 2\pi \sqrt{\delta}\,\, \mathcal{F}_\text{BH}\!\left[L_t/\sqrt{\delta},L_x/\sqrt{\delta}\right],\qquad Q(L_t,L_x,K) = \sqrt{\delta}\,\,q\!\left[L_t/\sqrt{\delta},L_x/\sqrt{\delta}\right]
\label{entbh}
\end{align}
where we define $\delta=KN^3/3$. The functions $\mathcal{F}_\text{BH}[u,v]$ and $q[u,v]$ are determined as follows. One can show that for our solutions, $u,v$ are such that the quartic equation
\begin{align}
  s^4 = 2(s+i(u+v)/2)(s+i(u-v)/2)
\end{align}
for complex variable $s$ has precisely one root strictly lying within each quadrant of the complex plane. Let $s_\star$ denote the unique root with positive real part and negative imaginary part. Then,
\begin{align}
  s_\star = \mathcal{F}_\text{BH}[u,v] - iq[u,v]. 
  \label{eq: s BH parts}
\end{align}
In the simplifying case of equal angular momenta where $J_1=J_2=J$, the black hole entropy can be written explicitly as,  
\begin{eqnarray}
\mathcal{S}_{\rm BH}(L,0, K ) & = & 2\pi \sqrt{K} \sqrt{\frac{N^{3}}{12}}\,\left[ \sqrt{1+\sqrt{1+\frac{6L^{2}}{KN^{3}}}}-\sqrt{2}\right]
\label{ppent}
\end{eqnarray}
with $L=L_{t}=2(J+Q)$. Although a more precise derivation will be given in \cite{GravityFollowUp} (see also Appendix C.4), we note that these results can be obtained as a simple scaling limit of the corresponding relations for Chow's solutions (\ref{eq: Chow entropy},\ref{s1}) in which $J_{3}\rightarrow \infty$ with the remaining charges scaling like $\sqrt{J_{3}}$. 
\paragraph{}
Before comparing with the boundary theory, we must determine the range of parameters and charges for which we can reliably approximate M-theory by supergravity. This is only the case if the curvature is small everywhere in units of the Planck length. We must also ensure that the radius of the compact $x^{+}$ direction is large enough to avoid the presence of additional light states. As noted above, as we approach the conformal boundary, the invariant length of the compact circle shrinks towards zero, and thus any classical description breaks down. Nonetheless, we can still seek conditions under which the classical description makes sense within the bulk, and in particular near the black hole horizon.  
We find that these conditions are satisfied for $K>>N^{7/3}>>1$, with the charges $J_{1}$, $J_{2}$ and $Q$ growing like $\sqrt{KN^{3}}$.  
In particular,  this means that we may hold fixed the arguments $L_t/\sqrt{\delta}$ and $L_{x}/\sqrt{\delta}$ appearing in (\ref{entbh}) as we take $K,N$ large. 
\paragraph{}


\section*{Black hole entropy from quantum mechanics}
The holographic duality proposed above suggest that the degeneracy of BPS states in superconformal quantum mechanics with fixed values of the charges $L_{t}$, $L_{x}$ (and $L_{y}=0$) should be related to the black hole entropy provided we work inside the regime of validity of the gravitational description. In the following we will test this by computing the asymptotics of the index coefficients $\mathcal{C}[L_{t}, L_{x}, L_{y}, K]$ in the limit $K\rightarrow \infty$. More precisely we will calculate the asymptotic growth of the index coefficients for fixed $N>>1$ in the $K\rightarrow \infty$ with the remaining charges $J_{1}$, $J_{2}$, $Q_{1}$, $Q_{2}$ scaling like $\sqrt{K}$ (or, equivalently, $L_{t}$, $L_{x}$, $L_{y}\sim\sqrt{K}$). As expressed in (\ref{cvsdintro}), these integer coefficients correspond to alternating sums over degeneracies of BPS states with different values of $F=-2Q_{2}$.
\paragraph{}
In determining the asymptotics of the index coefficients we will adapt a standard strategy of analytic combinatorics which originates Hardy and Ramanujan's classic analysis \cite{HardyRamanujan} of the growth of partition numbers. The first step is to represent the coefficients as an appropriate contour integral, 
 \begin{eqnarray}
\mathcal{C}[L_{t},L_{x},L_{y},K] & = & \frac{1}{(2 \pi i)^{4}}\, \oint_{C_{\beta}} \, \frac{dq_{\tau}}{q_{\tau}^{K+1}}\,\oint_{C_{+}} \, \frac{dt}{t^{L_{t}+1}}\,
\oint_{C_{-}} \, \frac{dx}{x^{L_{x}+1}}\,\oint_{C_{m}} \, \frac{dy}{y^{L_{y}+1}}\,
\,\, \mathcal{I}_{\rm QM}\left[t,x,y, q_{\tau} \right] \nonumber \\
\label{int1intro}
\end{eqnarray}
with contours specified in Appendix D.5 below. Continuing with the standard strategy, the next step is to determine the analytic properties of the integrand. In particular, we will 
show that the integrand has an essential singularity in $\beta=-\log q_{\tau}$ at $\beta=0$. Holding the remaining parameters, $t=\exp(-\epsilon_{+})$, $x=\exp(-\epsilon_{-})$ and $y=\exp(-m)$, at fixed complex values with ${\rm Im}[\epsilon_{+}]<0$ in the scaling limit, the relevant asymptotic of the index derived in our analysis is\footnote{We also find asymptotics for ${\rm Im}[\epsilon_{+}]>0$  which are related to (\ref{asymp}) by $\epsilon_{1}$, $\epsilon_{2}\rightarrow -\epsilon_{1}$, $-\epsilon_{2}$.}
\begin{eqnarray}
\log\,\mathcal{I}_{QM} & \begin{array}{c}  _{\beta\rightarrow 0} \\ \sim  \\ \,\,\end{array} & -\frac{N^{3}}{24}\,\frac{\Delta_{1}^{2}\Delta_{2}^{2}}{\epsilon_{1}\epsilon_{2}\beta} \,\,\,+\,\,\,O\left(\beta^{0}\right) 
\label{asymp}
\end{eqnarray} 
where $\epsilon_{\pm}=(\epsilon_{1}\pm \epsilon_{2})/2$, $m=(\Delta_{1}-\Delta_{2})/2+i\pi $   and,  
\begin{eqnarray}
\Delta_{1}+\Delta_{2}-\epsilon_{1}-\epsilon_{2} & = & 2\pi i 
\nonumber 
\end{eqnarray}
More precisely, these asymptotics hold up to parametrically small corrections if\footnote{Our result in the latter case of small $\epsilon_{1}$, $\epsilon_{2}$ agrees with that of \cite{LeeNahmgoong}, although their result is derived for real values of the parameters $\epsilon_{1}$ and $\epsilon_{2}$ with opposite sign.} {\em either} $N>>1$ {\em or} $|\epsilon_{1}|$, $|\epsilon_{2}|<<1$. Similar asymptotics appear in the large-$N$ and Cardy limits for indices of field theories in higher dimensions where they are related to the supersymmetric Casimir energy 
\cite{Casimir} and also the the entropy of asymptotically AdS black holes\footnote{Indeed such a relation exists \cite{Hosseini:2018dob, Choi:2018hmj} between the superconformal index of the $(2,0)$ theory itself and the Chow black hole solutions discussed above which is clearly related to the match found here.} \cite{HHZ}.  
\paragraph{}
Our study is aided by the coincidence noted above between the superconformal index and the much studied Nekrasov partition function \cite{Nekr} of $U(N)$ supersymmetric Yang-Mills theory on $\mathbb{R}^{4}\times S^{1}$ with a massive adjoint hypermultiplet. The index defined above can be written as, 
\begin{eqnarray}
\mathcal{I}_{\rm QM}\left[t,x,y,q_{\tau}\right] := 
\oint d\mu_{\rm Haar}\left[{Z}\right]\, \mathcal{Z}^{U(N)}_{\rm inst}\left[t,x,y,q_{\tau};{Z}\right] 
\label{indnekintro}
\end{eqnarray}
Here $\mathcal{Z}^{U(N)}_{\rm inst}$ is the non-perturbative part of the Nekrasov partition $\mathcal{Z}_{\rm Nek}$ (see Appendix D.1 for a full definition), which is naturally defined as an infinite series of instanton contributions. However our main goal in this section is to understand the behaviour of the index in the limit $\beta=-\log q_{\tau} \rightarrow 0$, when the contributions from large instanton number become unsupressed. 
In order to make progress, we need to find an effective way of resumming this series to obtain analytic function of the fugacity $q_{\tau}=\exp(-\beta)$. 
For fixed values of the other parameters, we expect the series to have a finite radius of convergence in $q_{\tau}$. The sum then defines $\mathcal{Z}^{U(N)}_{\rm Nek}$ as a holomorphic function within this radius which can then be analytically continued to whole of the disk $|q_{\tau}|< 1$. Indeed this has been proven rigorously for some special values of the parameters \cite{Felder}.
An important special case is provided by the $U(1)$ theory where the instanton moduli space coincides with the Hilbert scheme of points on $\mathbb{C}^{2}$. In this case only, one finds a completely explicit resummation of the instanton series in Plethystic form 
\cite{carlsson,Warnaar}. 
In particular, for $|q_{\tau}|$, $|q_{1}|$, $|q_{2}|<1$, we have, 
\begin{eqnarray}
\log \mathcal{Z}^{U(1)}_{\rm inst}\left[t,x,y,q_{\tau}\right] & = &  \sum_{n=1}^{\infty} \,\frac{1}{n}\, \frac{q_{\tau}^{n}}{1-q_{\tau}^{n}} \,\frac{t^{n}}{y^{n}}\, 
\frac{\left(1-y^{n}q_{1}^{n}/t^{n}\right)\left(1-y^{n}q_{2}^{n}/t^{n}\right)}{
\left(1-q_{1}^{n}\right)\left(1-q_{2}^{n}\right)} 
\label{u1sum}
\end{eqnarray}
Provided $|q_{\tau}|< {\rm min}\{|yt^{-1}|, |y^{-1}t^{-1}|\}$, this sum is convergent and defines a holomorphic function which can be extended by analytic continuation to the whole of the domain, 
\begin{eqnarray}
D^{U(1)} & := & \left\{ q_{1},q_{2},q_{\tau},y \in \mathbb{C}^{\star}\,: |q_{1}|, |q_{2}|, 
|q_{\tau}|\,\,<1\right\} 
\label{du1}
\end{eqnarray} 
The series (\ref{u1sum}) for $\log\mathcal{Z}^{U(1)}_{\rm inst}$ easily yields the asymptotic \cite{KimNahmgoong},  
\begin{eqnarray}
\log \mathcal{Z}^{U(1)}_{\rm inst}\left[t,x,y,q_{\tau}\right] & \sim & -\frac{1}{24}
\frac{m^{2}(2\pi i-m)^{2}}{\epsilon_{1}\epsilon_{2}\beta} 
\label{u1result}
\end{eqnarray}
for $|\epsilon_{1}|$, $|\epsilon_{2}|<<1$. These asymptotics allow for a rigorous saddle-point analysis of the resulting integral (\ref{int1intro}) which will be given in \cite{Andypaper}.  
\paragraph{}
As in the original analysis of Hardy and Ramanujan, the key step in the general case $N\geq 1$ is to exploit the modular properties of the integrand to determine its behaviour near the singular point. Our analysis follows closely a related discussion in \cite{LeeNahmgoong}.  
The modular properties of the Nekrasov partition function have a clear physical origin which can be understood by embedding the 5d gauge theory in string/M-theory. As discussed at various points in this paper, maximally supersymmetric five dimensional $U(N)$ gauge theory arises from the the $S^{1}$ compactification of the $(2,0)$ theory of type $A_{N-1}$. The latter is the worldvolume theory of $N$ M5 branes in M-theory. The Nekrasov partition function is itself an index  which counts BPS states of this system or, equivalently, a twisted Euclidean partition function for the $(2,0)$ CFT on $T^{2}\times \mathbb{R}^{4}$ where the $T^{2}$ is a torus of complex structure $\tau$ with an associated $SL(2,\mathbb{Z})$ group of modular transformations.  
\paragraph{}
The usual definition of the Nekrasov partition function as an infinite series of instanton corrections comes from a limit where the M-theory circle is small and M5 branes reduce to D4 branes in the Type IIA theory. In this picture instantons correspond to D0 branes and the $K$-instanton contribution counts BPS boundstates of $K$    
D0 branes and $N$ D4 branes. However, the index is deformation invariant and can also be computed in an M-theory regime where the relevant degrees of freedom are M2 branes stretched between M5 branes or "M-strings" \cite{mstrings}. In this regime the same partition function can be evaluated by calculating the elliptic genus of the M-string worldsheet theory in each charge sector. The resulting expression is, 
\begin{eqnarray}
\mathcal{Z}^{U(N)}_{\rm ell} & := & \left[\mathcal{Z}^{U(1)}_{\rm Nek}\right]^{N}\,Z_{S} 
\nonumber 
\end{eqnarray}
Where the M-string partition function $Z_{S}=Z_{S}[\epsilon_{1},\epsilon_{2},m,\tau;\mathbf{v}]$ is written as a function of the chemical potentials. To avoid clutter in our formulae, we will often suppress dependence on the remaining parameters and write $Z_{S}=Z_{S}[\mathbf{v}]$. Here $Z_{S}[\mathbf{v}]$ is defined as an expansion in powers of the $SU(N)$ fugacities $W_{a}:=\exp(-v_{a})$, for $a=1,2,\ldots,N-1$. 
\begin{eqnarray}
Z_{S}[\mathbf{v}] & := & \sum_{\mathbf{n}\in \mathbb{Z}_{\geq 0}^{N-1}}\, Z^{(\mathbf{n})}_{S}\, 
\exp\left(-\mathbf{n\cdot v}\right) 
\label{vevseries}
\end{eqnarray}
\paragraph{}
The coefficents in this expansion come from a localisation computation of the M-string worldsheet elliptic genus \cite{ell3, ell1, ell2} and are given explicitly in Appendix D.2. In this expression the instanton expansion has been resummed at each order in $w_{a}=\exp(-v_{a})$, to give an analytic function of the complex parameter $\tau=-2\pi i\beta$ 
defined for ${\rm Im}\tau>0$, which is elliptic in the parameters $\epsilon_{1}$, $\epsilon_{2}$ and $m$. The resulting coefficient functions are manifestly modular under $SL(2,\mathbb{Z})$ transformations acting on $\tau$. 
In particular, the $S$-generator of the modular group maps $\tau$ to $\tilde{\tau}:=-1/\tau$ and acts on the other parameters of the partition function as, 
\begin{eqnarray}
\epsilon_{1}\rightarrow \tilde{\epsilon}_{1}:=\epsilon_{1}/\tau \qquad{} & \qquad 
\epsilon_{2}\rightarrow \tilde{\epsilon}_{2}:=\epsilon_{2}/\tau \qquad{} & \qquad  
m \rightarrow \tilde{m}:=m/\tau 
\label{rep}
\end{eqnarray}
In the following we will use a "$\,\,\tilde{}\,\,$" to denote this replacement in functions of the parameters $\{\epsilon_{1},\epsilon_{2},m,\tau\}$. The modular transformation property for the coefficents $Z_{S}^{(\mathbf{n})}$ is \cite{KimNahmgoong}, 
\begin{eqnarray}
Z_{S}^{(\mathbf{n})} &= & \mathcal{K}_{+}^{(\mathbf{n})}\,\tilde{Z}_{S}^{(\mathbf{n})}
\label{key}
\end{eqnarray}
with 
\begin{eqnarray}
\mathcal{K}^{(\mathbf{n})}_{+} & = & \exp\left[ \frac{\epsilon_{1}\epsilon_{2}}{2\beta}\, 
\sum_{a,b=1}^{N-1}\, n_{a}\Omega_{a,b}n_{b}\,\,+\,\,\frac{1}{\beta}\left(m^{2}-\epsilon_{+}^{2}\right)\,\sum_{a=1}^{N-1}\,n_{a}\right]
\label{kernel}
\end{eqnarray}
where $\Omega_{a,b}=2\delta_{a,b}-\delta_{|a-b|,1}$ is the $A_{N-1}$ Cartan matrix. The exponent in this expression is proportional to the central charge of the M-string worldsheet theory.  
\paragraph{}
The physical discussion given above strongly suggests that that the two different representations of the  
M5-brane partition function described above correspond to different expansions of the same analytic function,  
\begin{eqnarray}
\mathcal{Z}^{U(N)} & := & \mathcal{Z}^{U(N)}_{\rm Nek} =\mathcal{Z}^{U(N)}_{\rm ell}
\label{zeq}
\end{eqnarray}
The same assumption is implicit in the modular bootstrap approach \cite{Bootstrap, Duan1,  Duan2} to determining refined BPS invariants of elliptically-fibred Calabi-Yau manifolds and its predictions have been tested in this context.  
The equality of $\mathcal{Z}^{U(N)}_{\rm Nek}$ and $\mathcal{Z}^{U(N)}_{\rm ell}$ is also closely related to the invariance of the refined topological vertex formalism \cite{refined} under a change of preferred direction \cite{mstrings}. We have checked this equality for $N=2$ by expanding both sides in $q_{\tau}$ and $\exp(-v_{a})$ at the first few orders, but we do not know of a general proof. 
\paragraph{}
Through the relation (\ref{zeq}) the Nekrasov partition function inherits the modular properties of $Z_{\rm ell}$. In the present context, the $S$-transformation discussed above relates the desired asymptotics in the strong-coupling limit $\beta\rightarrow 0$ to 
the corresponding behaviour in the weak-coupling limit $\beta\rightarrow \infty$. In Appendix D.3/D.4, we exploit this modular transformation property to derive the asymptotic form  (\ref{asymp}). 
Using these results, we find that the dominant contribution to the integral (\ref{int1intro}) in the scaling limit of large $K$ comes from the region near $\beta=0$ allowing its evaluation by steepest descent. We rewrite the integral in the form\footnote{The parameters $\mu_{\beta}\ldots$ appearing in the limits of integration are related to the choice of contour in (\ref{int1intro}).},  
\begin{eqnarray}
\mathcal{C}(L_{t},L_{x},L_{y},K) & = & \frac{1}{(2 \pi i)^{4}}\, \int_{\mu_{\beta}}^{\mu_{\beta}+2\pi i} \, 
d\beta \, \int_{\mu_{+}}^{\mu_{+}+2\pi i} \, d\epsilon_{+} \, \int_{\mu_{-}}^{\mu_{-}+2\pi i} \, d\epsilon_{-}\, \int_{\mu_{m}}^{2\pi i +\mu_{m}} \, dm\,
\,\, \exp\left(\mathcal{S}\right) \nonumber \\
\label{int2intro}
\end{eqnarray}
where the exponent is given as,  
\begin{eqnarray}
\mathcal{S}+2\pi i Q_{2} & = & -\frac{N^{3}}{24}\,\frac{\Delta_{1}^{2}\Delta_{2}^{2}}{\epsilon_{1}\epsilon_{2}\beta}\,+\,\Delta_{1} Q_{1}\,+\, \Delta_{2} Q_{2}\,+\,\epsilon_{1} J_{1}\,+\,\epsilon_{2} J_{2}\,+\,\beta K \nonumber \end{eqnarray}
We find that the integral is dominated by the contribution of a complex conjugate pair of saddle-points at $\mathcal{S}=2\pi \sqrt{\delta} s_\star$ and $2\pi \sqrt{\delta} \bar{s}_\star$, where $\delta=KN^{3}/3$ as above, and $s_\star$ is the unique root with positive real part and negative imaginary part of the quartic equation, 
\begin{eqnarray}
s^{2}(s-iw)^{2} & = & 2\left(s+i(u+v)/2\right)\left(s+i(u-v)/2\right)
\label{final}
\end{eqnarray}
Hence, we find the exponential growth, 
\begin{eqnarray}
\log\, \left| \mathcal{C}[L_{t},L_{x},L_{y}, K]\right| & \sim & 2\pi \sqrt{\delta} 
\mathcal{F}_{\rm QM}\left[L_{t}/\sqrt{\delta},L_{x}/\sqrt{\delta},L_{y}/\sqrt{\delta}\right] 
\label{growth1}
\end{eqnarray}
and $\mathcal{F}_{\rm QM}[u,v,w]=\text{Re}\, s_\star$. In the case $w=0$ applicable to comparison with our gravity, we find an exact match with the function which determines the black hole entropy. 
\begin{eqnarray}
\mathcal{F}_{\rm QM}[u,v,0] & = &  \mathcal{F}_{\rm BH}[u,v]
\nonumber 
\end{eqnarray}
\paragraph{}
It follows that the growth of the index coefficients in the case $L_{y}=0$ is determined by the entropy of the corresponding supersymmetric $X$ black hole,
\begin{eqnarray}  
\log \, \left|\mathcal{C}[L_t,L_x,0,K]\right|  & \sim & \mathcal{S}_{\rm BH}(L_{t},L_{x},K) 
\nonumber 
\end{eqnarray}
This agreement is clearly in line with the "central dogma" \cite{Maldrev} which suggests that the black hole should contribute a factor $\exp\,\mathcal{S}_{\rm BH}$ to the partition function of the gravitational theory. It is striking that, for each value of the charges $L_{t}$, $L_{x}$ and $K$, the full exponential growth of the index is exactly captured by the unique supersymmetric black hole solution which obeys the non-linear constraint $Q=Q(L_t,L_x,K)$ described above\footnote{Of course we cannot rule out the presence of additional supersymmetric black holes which do not obey the non-linear contraint like those found in the $AdS_{5}$ case in \cite{Jorge}. However there are no known solutions of this type in $AdS_{7}$ which are relevant for the boundary Penrose limit considered here. If additional solutions do exist in our case, then their contribution to the growth of the index must be subleading.}. More speculatively, we can extend the comparison to include the phase of $\mathcal{C}$. Keeping only the leading-order contribution at each saddle point we find, 
\begin{eqnarray}
  \mathcal{C}[L_t,L_{x},0,K] & \sim  & 2  \cos\left(2\pi Q(L_t,L_x,K)\right)\,\exp\,\mathcal{S}_{\rm BH}(L_{t},L_{x},K)
\nonumber 
\end{eqnarray}
Imposing the quantization condition $Q\in \frac{1}{2}\mathbb{Z}$  we find the sign $\cos(2\pi Q)=(-1)^{-2Q}=(-1)^{F}$ in agreement with the expected contribution of the black hole microstates to the index.  

\paragraph{}
Finally, note that a version of this analysis, valid for $L_{t}^{2}>>N^{3}K/6$, holds for all values of $N$. This regime is analgous to the Cardy limit of higher dimensional CFTs and resulting picture is very similar to recent studies of the superconformal index of $\mathcal{N}=4$ SUSY Yang-Mills in this context\cite{sixteenth1,sixteenth2}.    
The complex conjugate pair of saddle-point actions can then be written as $\mathcal{S}(L) \pm i  
\mathcal{S}(L)$ with, 
\begin{eqnarray}
\mathcal{S}(L) & = & 2\pi\,\left(\frac{N^{3}}{24}L^{2}K\right)^{\frac{1}{4}}
\label{sl}
\end{eqnarray}
Adding together the contribution of both saddle points, and including the prefactor from small fluctuations around the stationary point, 
we obtain a formula for the asymptotic growth of the index coefficients in the Cardy regime, 
\begin{eqnarray}
\mathcal{C}\left(L_,0,0,K\right) & \sim & 
\frac{3\sqrt{2}}{16\pi}\,\frac{\mathcal{S}(L)}{L^{2}K}\,\exp\left[\mathcal{S}(L)\right] \cos\left(  
\mathcal{S}(L)+\frac{\pi}{4} \right)
\nonumber 
\end{eqnarray}
with $\mathcal{S}(L)$ given by (\ref{sl}). In the Cardy regime, this formula 
holds for all $N$. In the case $N=1$, we have tested this formula against a direct expansion of the localisation formula on a computer up to $K\sim L\sim 60$, finding good agreement\footnote{In particular the formula agrees the data in Appendix B up to errors of around $1$-$2\%$ in the exponent. The accuracy is in agreement with error estimates for subleading terms.}.  
\paragraph{}
We would like to thank Dario Martelli, Jorge Santos and Chiung Hwang for useful discussions. We would also like to thank Sam Crew for collaboration in the early stages 
of this project. This work has been partially supported by STFC consolidated grant ST/T000694/1.

\appendix
\section*{Appendices}

\section{Model and Index details}


\subsection{Defining the model}

The simplest description of the model considered in this paper is provided by embedding it in gauged supersymmetric matrix quantum mechanics, which can in turn be obtained by dimensional reduction from supersymmetric gauge theory in higher dimension. An efficent 
starting point is to consider $\mathcal{N}=4$ supersymmetric $U(K)$ gauge theory in three spacetime dimensions with a particular choice of matter content. In addition to the $\mathcal{N}=4$ 
vector multiplet, the theory contains an adjoint hypermultiplet whose lowest components are two complex  $K\times K$ matrix-valued scalar fields $X$ and $\tilde{X}$. We also include $N$ additional hypermultiplets in the fundamental representation of $U(K)$. 
The lowest components of these multiplets are scalar fields $Q_{i}$, $\tilde{Q}_{i}$ which are $K$-component complex vectors and also carry an aditional flavour index  $i=1,2,\ldots,N$ together with a $U(K)$ gauge field. By writing the theory in terms of complex fields, we are picking an $\mathcal{N}=2$ subalgebra of $\mathcal{N}=4$ supersymmetry in three dimensions. With respect to the chosen $\mathcal{N}=2$ subalgebra, each complex scalar field is the lowest component of a chiral multiplet. Apart from the 3d gauge coupling, the parameters of the theory are a triplet of masses for each hypermultiplet and a triplet of real FI parameters. In the $\mathcal{N}=2$ language it is natural to split each of these parameters into a single complex parameter and a real parameter.    
\paragraph{}
In the case of vanishing masses, the three-dimensional theory has a Higgs branch\footnote{The 3d theory also has a Coulomb branch which is not relevant to our discussion.} which is defined by solving the F- and D-term equations. In case of vanishing FI parameters, these take the form 
\begin{eqnarray}
\left[X,\tilde{X}\right] \,+\, \sum_{i=1}^{N}Q_{i}\tilde{Q}_{i} & = & 0  \nonumber \\ 
\left[X,X^{\dagger}\right] + \left[\tilde{X},\tilde{X}^{\dagger}\right] \,+\, \sum_{i=1}^{N} \, |Q_{i}|^{2}-|\tilde{Q}_{i}|^{2} & = & 0
\label{adhm}
\end{eqnarray} 
To find the space of inequivalent vacua we must also mod out by the action of the $U(K)$ gauge group. These steps coincide precisely with the definition of the moduli space $\mathcal{M}_{K,N}$ appearing in the ADHM contruction \cite{ADHM} of $K$ instantons of an $SU(N)$ gauge theory defined on $\mathbb{R}^{4}$. This is a singular space of complex dimension $2KN$. 
Turning on the FI parameters introduces non-zero constant terms on the RHS of the F- and D-term equations. This replaces the singular space $\mathcal{M}_{K,N}$ by a resolved space $\tilde{\mathcal{M}}_{K,N}$ which is a smooth hyper-K\"{a}hler manifold of the same dimension. Both the singular space and its resolution also have descriptions in algebraic geometry as Nakajima quiver varieties. In the abelian case $N=1$, this coincides with the Hilbert scheme of $K$ points on $\mathbb{C}^{2}$.          
\paragraph{}
The next step is to dimensionally reduce the three-dimensional gauge theory described above to obtain a theory in $(0+1)$ dimensions. The resulting quantum mechanics has a gauge coupling $e^{2}$ of mass dimension $3$. At energies high compared to the mass scale $\Lambda=e^{\frac{2}{3}}$ the theory is weakly coupled. At energies much lower than $\Lambda$, gauge interactions become strong and the theory flows to a non-linear $\sigma$-model with target $\mathcal{M}_{K,N}$. As the target space is hyper-K\"{a}hler, the $\sigma$-model has $\mathcal{N}=(4,4)$ supersymmetry\footnote{Strictly speaking this refers to the supersymmetry of the corresponding $\sigma$-model in $(1+1)$ dimensions but we refer here to its dimensional reduction to QM.}. The quantum mechanical gauge theory theory also has FI parameters and masses inherited from the corresponding parameters of the three dimensional theory. Introducing non-zero values for these parameters in the UV results in supersymmetry preserving deformations of the IR $\sigma$-model. As already mentioned turning on FI parameters replaces the singular target space $\mathcal{M}_{K,N}$ 
by its smooth resolution $\tilde{\mathcal{M}}_{K,N}$. Mass parameters in the UV theory are associated to global symmetries which are realised isometries of the target manifold in the IR. Turning on non-zero masses in the UV introduces a potential in the IR $\sigma$-model proportional to the norm of the corresponding Killing vector.      
\paragraph{}
In the following, we will be mainly interested in the IR limit of the matrix model described above. This corresponds to focussing on energies far below the mass scale set by gauge coupling or simply taking the limit $e^{2}\rightarrow \infty$. The resulting theory is a $(0+1)$-dimensional non-linear $\sigma$-model with target space $\mathcal{M}_{K,N}$. 
The singular space $\mathcal{M}_{K,N}$ is a hyper-K\"{a}hler cone. In other words, it admits a triholomorphic homothety. This condition ensures that the Poincare supersymmetry of the $\sigma$-model is enhanced to full superconformal invariance. The corresponding superconformal algebra is the simple Lie super-algebra $\mathfrak{g}:=\mathfrak{osp}(4^{*}|4)$ \cite{Singleton1}. Thus matrix quantum mechanics flows in the IR to a one dimensional superconformal theory with this symmetry. In analogy with the more familiar case of higher dimensional SCFT, we expect to find a spectrum of states transforming in irreducible, unitary representations of $\mathfrak{g}:=\mathfrak{osp}(4^{*}|4)$. \\
\paragraph{}
The maximal bosonic subalgebra of $\mathfrak{g}:=\mathfrak{osp}(4^{*}|4)$ is, 
\begin{eqnarray}
\mathfrak{g}_{B} & = & \mathfrak{so}(2,1)\oplus \mathfrak{su}(2) \oplus \mathfrak{so}(5)         
\label{bose}
\end{eqnarray}        
The first factor is the quantum mechanical superconformal algebra with generators 
$\{H,\Sp,D\}$ obeying, 
\begin{eqnarray}
[D, H]\,=\,2i H & \qquad{} [D, \Sp]\,=\,-2 i\Sp \qquad{} & [\Sp, H]\,=\,- iD  
\end{eqnarray}
The remaining factors in the bosonic subalgebra $\mathfrak{g}_{B}$, correspond to   
R-symmetries which act on the bosonic and fermionic factors respectively. 
The generators of the Lie superalgebra are completed by eight "Poincare" supercharges $\Q_{A}$ of positive dimension and eight superconformal charges $\S_{A}$ of negative dimension.  
\paragraph{}
As usual for supersymmetric quantum mechanics, the wavefunctions correspond to differential forms on the target manifold and the Hamiltonian $H$ is naturally identified with the corresponding Laplacian operator. The action of the full superconformal algebra on differential forms is constructed in \cite{Singleton1}. 
In this geometrical picture, the dilatation operator $D$ coincides with the Lie derivative corresponding to the homothetic Killing vector \cite{Michelson} while the special conformal generator $\Sp$ corresponds to a harmonic potential on the moduli space. The ADHM construction corresponds to a hyper-K\"{a}hler quotient construction of the moduli-space and one can define an action of $\mathfrak{so}(2,1)$ on the pre-quotient space spanned by the ADHM matrices which descends through the quotient to $\mathcal{M}_{K,N}$. In this context, the special conformal generator takes the form, 
\begin{eqnarray}
\Sp & = & |X|^{2}\,+\,|\tilde{X}|^{2}\,+\,\sum_{i=1}^{N}\,\left( |Q_{i}|^{2}\,+\,|\tilde{Q}_{i}|^{2} \right) \nonumber 
\end{eqnarray}
As the instanton moduli space is non-compact, the spectrum of $H$ is continuous when evaluated on plane wave normalisable states. However, by a standard argument, the operator $-iD$ has a discrete spectrum which coincides with that of the \textit{oscillator Hamiltonian} $\Hosc:=H+\Sp$.
\paragraph{}
Any non-flat hyper-K\"{a}hler cone is necessarily singular and hence the discussion given above is not precise in the general case. As mentioned above, 
the instanton moduli space $\mathcal{M}_{K,N}$ has a smooth resolution $\tilde{\mathcal{M}}_{K,N}$ induced by turning on FI parameters in the underlying $U(K)$ gauge theory. Here, in our chosen complex structure, we will only consider the real FI parameter 
$\xi_{\mathbb{R}}$ which deforms the D-term equation in (\ref{adhm})\footnote{In this complex structure, the resolved moduli-space $\tilde{\mathcal{M}}_{K,N}$ is a holomorphic symplectic manifold known as the symplectic resolution of the singular space $\mathcal{M}_{K,N}$.}. 
This deformation breaks conformal symmetry while preserving the usual Poincare supersymmetry of the $\sigma$-model. In particular, as the resolved space admits a smooth hyper-K\"{a}hler metric, $\mathcal{N}=(4,4)$  supersymmetry remains unbroken. The $\mathfrak{su}(2)$ factor in the bosonic subalgebra $\mathfrak{g}_{B}$ is broken to a $\mathfrak{u}(1)$ which corresponds to a holomorphic isometry of the resolved space. 
The harmonic potential $\Sp$, corresponds to the norm of the corresponding Killing vector and can therefore also be introduced in a way which preserves supersymmetry.   
With the FI parameter, $\xi_{\mathbb{R}}$ turned on, the model is well defined and the Hamiltonian as $\Hosc=H+\Sp$ has a discrete spectrum which is bounded below.
In the limit that the the resolution parameters go to zero, it is natural to expect that the corresponding states will lie in multiplets of the full superconformal algebra. 
Following \cite{ABS}, we will consider the resolution  
as UV regulator for the theory analogous to the usual situation in higher dimensional field theory.
\paragraph{}
In our discussion, the algebra $\mathfrak{osp}(4^{*}|4)$ arises in three different ways which are related by dualities I and II described in the introduction. The first is as the superconformal algebra of quantum mechanics as described above. The second is as the light-cone reduction of the superconformal algebra $\mathfrak{osp}(8^{*}|4)$ of the $(2,0)$ theory in six dimensions. Finally, the same algebra characterises the (super-)isometries of the dual gravitational background which corresponds to a certain null compactification of $AdS_{7}\times S^{4}$. We will now introduce some global conventions for the generators of the algebra and the corresponding chemical potentials/fugacities. \\

We define Cartan subgroup generators $\Hosc$, $J_{R}$ and $Q_{1}$, $Q_{2}$ 
for each factor in the maximal bosonic subgroup, 
\begin{equation}
SO(2,1) \times SU(2)_{R} \times SO(5)\subset OSp(4^{*}|4) 
\label{superconf}
\end{equation} 
of the superconformal group. 
Here the label $R$ distinguishes the $SU(2)$ factor in the superconformal group from the global symmetry $SU(2)_{L}$ of the instanton moduli space which commutes with all superconformal generators. The Cartan generator for the latter symmetry is denoted $J_{L}$. The $SU(2)_{L}\times SU(2)_{R}\simeq SO(4)$ symmetry originates in the rotational symmetry of the $\mathbb{R}^{4}$ on which the instantons are defined. Sometimes it will be convenient to work instead with the generators $J_{1}$ and $J_{2}$ of rotations in two orthogonal places of $\mathbb{R}^{4}$ in terms of which $J_{R}=J_{1}+J_{2}$, $J_{L}=J_{1}-J_{2}$. We also define similar combinations for the Cartan generators of $SO(5)$: $Q_{R}:=Q_{1}+Q_{2}$, $Q_{L}:=Q_{1}-Q_{2}$. 
For convenience, we will use the same notation for the generators and for the corresponding eigenvalues. We work in a normalisation such that $J_{1}$, $J_{2}$, $Q_{1}$ and $Q_{2}$ are, in general, half-integer valued. However, for all states which contribute to the index it turns out that the combinations $J_{L}$, $J_{R}$, $Q_{L}$, $Q_{R}$ defined above take integer values.   
\paragraph{}
In addition to the superconformal invariance described above the quantum mechanical model also has global symmetries corresponding to the isometries of the instanton moduli space. 
The resulting global symmetry group is, 
\begin{equation}
SU(2)_{L}\times SU(N)
\label{global}
\end{equation}
where $SU(2)_{L}$ corresponding to remaining factor in the rotation group on $\mathbb{R}^{4}$ and the $SU(N)$ factor originates from the Yang-Mills gauge symmetry. We introduce 
Cartan generators/eigenvalues $n_{a}$, for $a=1,2,\ldots N-1$. In the following we will also introduce chemical potentials $\epsilon_{-}=(\epsilon_{1}-\epsilon_{2})/2$ for the Cartan generator of $SU(2)_{L}$ and chemical potentials $v_{a}$ for generators $n_{a}$ of SU(N), with $a=1,2,\ldots,n$. We will typically use vector notation for the $SU(N)$ charges and chemical potentials., with  $\mathbf{n}:=(n_{1},n_{2},\ldots,n_{N-1})\in \mathbb{Z}^{N-1}$  and $\mathbf{v}:=(v_{1},v_{2},\ldots,v_{N-1})\in \mathbb{C}^{N-1}$.

\subsection{Localisation Formulae}

We establish the following conventions in relation to the localisation formulae for the superconformal index, the M-string elliptic genus and the Nekrasov partition function.

\paragraph{}
A partition $\lambda$ is a finite sequence of nonincreasing positive integers $\lambda_1\geq \lambda_2\geq \lambda_3... \geq \lambda_{\ell(\lambda)}>\lambda_{\ell(\lambda)+1}=0$. Here $\ell(\lambda)$ is called the length of the partition and $|\lambda|=\sum_{p}\lambda_{p}$ is its weight. The set of partitions is denoted $\mathcal{P}$. 
\paragraph{}
One usually visualizes a partition $\lambda$ by drawing the associated Young diagram $Y(\lambda)$. A box $s$ in a young diagram $Y(\lambda)$ is labelled by its cooordinates $(p,q)$ where $p = 1,..., l(\lambda)$. Arm and leg lengths of a box $s=(p,q)\in Y(\lambda)$  
are defined as,  

\begin{eqnarray}\label{leg_arm_length}
    a(s) = \lambda_p - q, & \qquad{} & l(s)= \lambda^{\vee}_q - p
\end{eqnarray}
where the dual $\lambda^{\vee}$ of a partition $\lambda$ is obtained by interchanging rows and columns in the Young diagram. This definition can extended to boxes $s=(p,q)$ which lie outside the Young diagram.
\paragraph{}
We also consider $N$-component vectors of partitions denoted $\vec{\lambda}=(\lambda^{(1)},\lambda^{(2)},\ldots,\lambda^{(N)})$ with total weight, 
\begin{eqnarray}
||\vec{\lambda}|| & = & \sum_{i=1}^{N}\, |\lambda^{(i)}|
\label{tot}
\end{eqnarray}
For each box $s\in Y(\lambda^{(i)})$ we define, 
\begin{eqnarray}
g_{ij}(s) & = & -a_{i}(s)+l_{j}(s) \nonumber \\ 
f_{ij}(s) & = & -a_{i}(s)-l_{j}(s)-1
\label{fg}
\end{eqnarray}
Here $a_{i}(s)$ and $l_{j}(s)$ are the arm and leg lengths of box $s$, relative to the Young Tableaux $Y(\lambda^{(i)})$ and $Y(\lambda^{(j)})$ respectively. 
\paragraph{}
We define a  Plethystic exponental for any function $f$ of $r$ formal variables $\{x_{1},x_{2},\ldots,x_{r}\}$ as, 
\begin{eqnarray}
{\rm Pexp}\left[f(x_{1},x_{2},\ldots,x_{r})\right) & = & 
\exp \left[\sum_{n\geq 1} \frac{1}{n} f\left(x^{n}_{1},x^{n}_{2},\ldots,x^{n}_{r}\right) \right]
\nonumber{} 
\end{eqnarray}
For any $X$ we also define, 
\begin{eqnarray}
\left[ X\right] & := & \sqrt{X}-\frac{1}{\sqrt{X}} 
\nonumber 
\end{eqnarray}
With these definitions the explicit formula for the superconformal index can be written in the compact form (\ref{loc1}). 

\section{Computational Results} 
Evaluation of the index coefficients $\mathcal{C}[L,0,0,K]$ for various values of $K$ and $L$ in the abelian case $N=1$. The second table focusses specifically on the case $K=L$.  
\begin{center}
\begin{tabular}{|c|c|}
\hline
$(L, K)$ & $\mathcal{C}(L,0,0,K)$\\
\hline
(24, 25) & -4548136426 \\
(26, 26) & 7935209206\\
(28, 27) & 88800402896\\
(30, 28) & -4887654890 \\
(32, 29) & -1425581403152\\
(40, 45) & 7337290205677620\\
(42, 46) & 23800263998384620\\
(44, 47) & -51625798313702826 \\
(46, 48) & -429211479407800616 \\
(48, 49) & -288354194415296772 \\
(50, 50) & 4773158006473089778 \\
(52, 51) & 14870285533157146362\\
(54, 52) & -21630735101481854366 \\
\hline
\end{tabular}
\end{center}

{\footnotesize
\begin{tabular}{c|c|c|c|c|c}
\multicolumn{2}{c}{}&\multicolumn{4}{c}{$L$}\\
\multicolumn{1}{c}{}&& 40 & 42 & 44 & 46\\
\hline
 &45 & 7337290205677620 & 22476503716465140 & 9610110542310490 & -143942357454816860\\
&46 & 10640057177590492 & 23800263998384620 &-14596846924592666 & -231844859695908664\\
K&47 & 14121688544604138 & 21607000602591350 & -51625798313702826 & -330802093776910478\\
&48 & 17342715906788200  & 14262589686082038 & -102521859977682242 & -429211479407800616\\
&49 & 19682978250039222 & 60266617136788 & -166700677839463202 & -509237288538994006\\
\hline
\end{tabular}
}

\begin{tabular}{c|c|c|c|c}
\multicolumn{2}{c}{}&\multicolumn{3}{c}{$L$}\\
\multicolumn{1}{c}{}&& 50 & 52 & 54\\
\hline
 &50 & 4773158006473089778 & 13722136087430823474 & 10728665632616173124 \\
&51 & 6943033937905529622 & 14870285533157146362 & -1493086283031736666 \\
&52 & 9190525712121239144 & 13822179413239343452 & -21630735101481854366 \\
&53 & 11174425419671147488 & 9447220680249748082 & -50744695186842694114 \\
K&54 & 12412734295210394496 & 543115812956557290 & -88938229446341455870 \\
&57 & 4868162317987965318 & -63009089217696929546 & -233897399529627516932 \\
&58 & -4069732338637176522 & -97265673400284395108 & -272448778843705818614 \\
&59 & -17508654288324952938 &-136216858133084019088 & -288990007308923696954\\
& 60 &-35960421557696977470 & -176744583922449466508 &-268821840220720476958\\
\hline
\end{tabular}

\section{Details of black hole solution}\label{app: BH solution}

In this Appendix, we present an explicit supergravity black hole solution realising the desired $\ncAdS\times S^4$ asymptotics, where recall, $\ncAdS$ is a particular null compactification of AdS$_7$. Following \cite{Chong:2004dy,Chow:2007ts}, we will proceed by seeking a solution to $\mathcal{N}=4$ $SO(5)$ gauged supergravity in seven dimensions, obtained by reduction of eleven-dimensional supergravity on $S^4$ \cite{Nastase:1999cb,Nastase:1999kf}, which is further truncated to include only a pair of Abelian gauge fields in the $U(1)^2\subset SO(5)$ Cartan subgroup. In addition to the metric and this pair $A_{(1)}^{1,2}$ of 1-form gauge fields, the bosonic field content also contains a pair of scalars $X_{1,2}$ and a 3-form $A_{(3)}$. There is also a gauge coupling $g$. Finally, there is additionally a self-duality condition imposed upon the fields, written in terms of an auxiliary 2-form $A_{(2)}$. We use a convenient formulation of the Lagrangian as first set out in \cite{Cvetic:2000ah}.
\paragraph{} 
Our solution has equal scalars $X=X_1=X_2$ and equal 1-forms $A_{(1)}=A_{(1)}^1 = A_{(1)}^2$. Writing then $F_{(2)}=dA_{(1)}$ and $F_{(4)}=dA_{(3)}$, the bosonic field equations are found by variation of the Lagrangian
\begin{align}
  \mathcal{L} &= R\star 1 - 5 X^{-2}dX \wedge \star dX - X^{-2} F_{(2)}\wedge \star F_{(2)} - \frac{1}{2}X^4 F_{(4)}\wedge \star F_{(4)}		\nn\\
  &\qquad + 2g^2 (8X^2+8X^{-3} - X^{-8} )\star 1 + F_{(2)}\wedge F_{(2)} \wedge A_{(3)} + g F_{(4)}\wedge A_{(3)}
  \label{eq: Lagrangian}
\end{align}
while we must also satisfy the self-duality equation
\begin{align}
  X^4 \star F_{(4)} = 2g A_{(3)} - F_{(3)} + F_{(2)}\wedge A_{(1)}
  \label{eq: self-duality equation}
\end{align}
where $F_{(3)}=dA_{(2)}$ is the field strength of some 2-form $A_{(2)}$.

\subsection{Statement of the solution}

Our solution is in terms of coordinates $(u,x^+,x^-,y,z,\phi_\alpha)$, $\alpha=1,2$, with $\phi_\alpha\sim \phi_\alpha + 2\pi$. In addition to the gauge coupling $g$, it depends on five parameters $(m,\delta,a_1,a_2,\lambda)$, where the correct signature requires $a_1 g, a_2 g\in (-1,1)$. Crucially, our solution realises $X$ asymptotics, given by the metric (\ref{eq: AdS7 pp-wave slicing}), where we see that the gauge coupling $g$ is identified as the inverse AdS radius $g=R_\text{AdS}^{-1}$ of the asymptotic geometry, and thus is further determined in terms of the rank $N$ as in (\ref{eq: AdS radius in Planck units}).
\paragraph{}
The metric is
\begin{align}
  ds^2 = H^{2/5} \Bigg[& \frac{(u^2 + y^2)(u^2 + z^2)}{U} du^2 + \frac{(u^2+y^2)(y^2 - z^2)}{Y}dy^2 + \frac{(u^2 + z^2)(z^2 - y^2)}{Z}dz^2	\nn\\
  &\quad - \frac{U}{H^2 (u^2+y^2)(u^2 + z^2)} \B[y^2,z^2,0]^2 		\nn\\
  &\quad + \frac{Y}{(u^2 + y^2)(y^2-z^2)} \left(\B[-u^2, z^2,0]- \frac{q}{H(u^2 + y^2)(u^2 + z^2)}\B[y^2,z^2,0]\right)^2	\nn\\
  &\quad + \frac{Z}{(u^2 + z^2)(z^2-y^2)} \left(\B[-u^2, y^2,0]- \frac{q}{H(u^2 + y^2)(u^2 + z^2)}\B[y^2,z^2,0]\right)^2	\nn\\
  &\quad + \frac{a_1^2 a_2^2 }{g^2u^2 y^2 z^2} \left(\B[-u^2,y^2,z^2] - \frac{q}{H(u^2 + y^2)(u^2 + z^2)}\left(1-\frac{g^2y^2 z^2}{a_1 a_2 }\right)\B[y^2,z^2,0]\right)^2 \,\,\Bigg]
  \label{eq: BH plane wave metric}
\end{align}
where
\begin{align}
  U(u) &= \frac{(1+g^2 u^2)^2}{g^2u^2} (u^2 + a_1^2)(u^2 + a_2^2) + qg^2 \left(2u^2 + a_1^2 + a_2^2 + \frac{1}{g^2}\right)  + \frac{2qa_1 a_2}{u^2} + \frac{q^2 g^2}{u^2} -2m		\nn\\
  Y(y) &= \frac{\left(1-g^2 y^2\right)^2}{g^2y^2} (a_1^2 - y^2)(a_2^2 - y^2),	\qquad Z(z) = \frac{\left(1-g^2 z^2\right)^2}{g^2 z^2} (a_1^2 - z^2)(a_2^2 - z^2)		\nn\\
  H(u,y,z) &= 1+ \frac{q}{(u^2 + y^2)(u^2 + z^2)},\quad q = 2m\sinh^2 \delta 
\end{align}
and, given any three arguments $v_1,v_2,v_3$, we define the one-form
\begin{align}
  &\B[v_1,v_2,v_3] \nn\\
  &= \frac{\lambda}{g }\frac{(1-g^2 v_1)(1-g^2 v_2)(1-g^2 v_3)}{(1- a_1^2 g^2)(1- a_2^2 g^2)}\,dx^+				\nn\\[0.5em]
  &\quad + 	\frac{1 }{2g} \frac{1}{(1- a_1^2 g^2)^2(1- a_2^2 g^2)^2}\bigg( 	(1-3 a_1^2 g^2 - 3 a_2^2 g^2 + 5 a_1^2a_2^2 g^4) \nn\\
  &\hspace{60mm} + g^2 (1 + a_1^2 g^2 + a_2^2 g^2 - 3 a_1^2 a_2^2 g^4)(v_1 + v_2 + v_3)				\nn\\
  & \hspace{60mm}+ g^4 (-3 + a_1^2 g^2 + a_2^2 g^2 + a_1^2 a_2^2 g^4)(v_1 v_2 + v_2 v_3 + v_3 v_1)			\nn\\
    & \hspace{60mm} +g^6 (5 - 3 a_1^2 g^2 - 3 a_2^2 g^2 + a_1^2 a_2^2 g^4)v_1 v_2 v_3 \bigg) dx^-			\nn\\[0.5em]
  &\quad + \frac{g^2(a_1^2-v_1)(a_1^2-v_2)(a_1^2-v_3)}{a_1(1- a_1^2 g^2)^2(a_1^2 - a_2^2)} d\phi_1 + \frac{g^2(a_2^2-v_1)(a_2^2-v_2)(a_2^2-v_3)}{a_2(1- a_2^2 g^2)^2(a_2^2 - a_1^2)} d\phi_2
  \label{eq: B def}
\end{align}
The solutions for the scalars and 1-forms are
\begin{align}
  X_1 = X_2 = H^{-1/5},\quad A_{(1)}^1 = A_{(1)}^2 = \frac{2m \sinh\delta \cosh\delta}{H(u^2 + y^2)(u^2 + z^2)}\B[y^2,z^2,0]
  \label{eq: scalars and 1-forms}
\end{align}
while for the 3-form we have
\begin{align}
  A_{(3)} &= -q g^3 a_1 a_2  \Big(\B[y^2,z^2,0] - \B[y^2,z^2,g^{-2}]\Big)\wedge \left(\frac{1}{(u^2+y^2)z}dz\wedge \Big(\B[y^2,0,0] - \B[y^2,0,g^{-2}]\Big)\right.		\nn\\
    &\hspace{80mm}+ \left.\frac{1}{(u^2+z^2)y}dy\wedge \Big(\B[z^2,0,0] - \B[z^2,0,g^{-2}]\Big)\right)		\nn\\
  & \quad + qg^3\B[y^2,z^2,0]\wedge \left(\frac{z}{(u^2+y^2)}dz\wedge \Big(\B[y^2,0,0] - \B[y^2,0,g^{-2}]\Big)\right.		\nn\\
    &\hspace{45mm}+ \left.\frac{y}{(u^2+z^2)}dy\wedge \Big(\B[z^2,0,0] - \B[z^2,0,g^{-2}]\Big)\right)		
\end{align}
The 2-form $A_{(2)}$ appearing in the self-duality equation (\ref{eq: self-duality equation}) can be taken as
\begin{align}
  A_{(2)} &= \frac{q}{H(u^2 + y^2)(u^2 + z^2)} \B[y^2, z^2,0]\wedge \Bigg((1-a_1 a_2 g^2)\B[y^2,z^2,g^{-2}]		\nn\\
  & \hspace{70mm}+ \frac{(a_1^2 - y^2)(a_1^2-z^2)}{a_1(a_1^2-a_2^2)}d\phi_1+ \frac{(a_2^2 - y^2)(a_2^2-z^2)}{a_2(a_2^2-a_1^2)}d\phi_2\Bigg)
  \label{eq: 2-form solution}
\end{align}
Our solution was derived in a particular parameter coordinate limit of the solution of Chow \cite{Chow:2007ts}, following the procedure put forward in \cite{Maldacena:2008wh}. The fact that the Chow configuration is indeed a supergravity solution implies that our solution is too; nonetheless, we have performed an independent check in \textit{Mathematica} that confirms that this is indeed the case.
\paragraph{}
For generic values of the parameters $(m,\delta,a_1,a_2,\lambda)$, the black hole (\ref{eq: BH plane wave metric}) is non-extremal and breaks all supersymmetry. However, we can preserve $\frac{1}{8}$ supersymmetry, if the parameters are constrained to satisfy
\begin{align}
  e^{2\delta} = 1+ \frac{2}{(1-a_1 g - a_2 g)}
  \label{eq: SUSY condition}
\end{align}
The black holes of \cite{Chow:2007ts}---of which our solutions are a Penrose limit---have an additional constraint amongst parameters, required in order to avoid the presence of naked closed timelike curves. In particular, this constraint admits a smooth Penrose limit, and thus we assume it must also be imposed upon our solutions. It has the effect of fixing the parameter $m$, as
\begin{align}
  q = 2m \sinh^2\delta = -\frac{2}{g^3}\frac{(a_1+a_2)(1-a_1 g)^2(1-a_2g)^2}{(2-a_1 g - a_2 g)^2}
  \label{eq: q constraint}
\end{align}
In line with much of the literature, we distinguish a black hole satisfying both of these conditions as a ``BPS" black hole. In particular, these two conditions are sufficient to ensure the vanishing of the temperature of the black hole; BPS black holes are extremal.

The event horizon of such a BPS black hole is a surface of constant $u=u_0$, where $u_0$ is the largest root of $U(u)$. Indeed, the extremality of the BPS solution corresponds to this root being a double root. We have explicitly,
\begin{align}
  u_0^2 = \frac{2g a_1 a_2 - a_1-a_2}{(2-a_1 g - a_2 g)g}
\end{align}
A real solution of the supersymmetry condition (\ref{eq: SUSY condition}) for the parameters $(m,\delta,a_1,a_2,\lambda)$ requires $e^{2\delta}>0$, while if we want a horizon we need $u_0^2 >0$. These two constraints are satisfied precisely if, in addition to the existing $a_1 g,a_2 g\in (-1,1)$, we have
\begin{align}
  2 (a_1 g) (a_2 g) - a_1 g-a_2g>0
  \label{eq: a1 a2 bounds}
\end{align}
Note, this condition then implies $g(a_1+a_2)<0$, which by (\ref{eq: q constraint}) ensures that $m>0$.

\subsection{Asymptotic geometry}

The metric (\ref{eq: BH plane wave metric}) realises the AdS$_7$ pp-wave slicing (\ref{eq: AdS7 pp-wave slicing}) asymptotically. More precisely, there exists coordinates $(r,x^+, x^-, n_\alpha,\phi_\alpha)$ such that as we approach the conformal boundary $r\to\infty$ (corresponding to $u\to\infty$), we have
\begin{align}
  ds^2\longrightarrow \frac{dr^2}{g^2r^2} + r^2 \left(-2dx^+ dx^- - \rho^2 (dx^-)^2 + \sum_\alpha \left(dn_\alpha^2 + n_\alpha^2 d\phi_\alpha^2\right)\right)
\end{align}
To get to this coordinate system, we exchange the coordinates $(u,y,z)$ with $(r,n_1,n_2)$ using the relations
\begin{align}
  	(1- a_1^2 g^2)^2 r^2 n_1^2 		&= -\frac{(a_1^2 - y^2)(a_1^2 - z^2)}{(a_1^2 - a_2^2)}g^2(u^2 + a_1^2) 	\nn\\
  	(1- a_2^2 g^2)^2 r^2 n_2^2 		&= -\frac{(a_2^2 - y^2)(a_2^2 - z^2)}{(a_2^2 - a_1^2)}g^2(u^2 + a_2^2) 	\nn\\
  	r^2								&= \frac{(1-g^2 y^2)(1-g^2 z^2)}{(1- a_1^2 g^2)(1- a_2^2 g^2)}\frac{\lambda}{g^2}(1+g^2 u^2)
  	\label{eq: new implicit relations}
\end{align}
Indeed, it is this implicit transformation that fixes the range of $(y,z)$. Assuming $a_1^2<a_2^2<g^{-2}$, we can take $y,z\ge 0$ with
\begin{align}
  a_1^2 \le y^2 \le a_2^2 \le z^2 \le g^{-2}
\end{align}
Clearly, the transformation (\ref{eq: new implicit relations}) breaks down if $a_1\to a_2$ and/or $a_2 \to g^{-1}$; indeed, the entire black hole solution (\ref{eq: BH plane wave metric}) becomes singular in either limit. However, one can remedy this by first making a further coordinate transformation, replacing $(y,z)$ with a pair of angles $(\theta_1,\theta_2)$ with $0\le \theta \le \pi/2$ and
\begin{align}
  y^2 = a_1^2 + (a_2^2-a_1^2)\sin^2 \theta_1,\qquad z^2 = a_2^2 + (g^{-2}-a_2^2)\sin^2 \theta_2
  \label{eq: y,z angle replacement}
\end{align}
Then the transformation (\ref{eq: new implicit relations}) is perfectly regular when either (or both) problematic limits are taken, and further the solution (\ref{eq: BH plane wave metric}) is regular in either or both limits when written in terms of $(u,x^+,x^-,\theta_1,\theta_2,\phi_1,\phi_2)$.
\paragraph{}
Thus, in order to define a black hole solution with $\ncAdS$ asymptotics, we simply identify the coordinate $x^+\sim x^+ + 2\pi$ throughout the spacetime.
\subsection{Length of the compact circle}

The norm of the vector field $\partial_+$ approaches zero as we approach the conformal boundary; indeed, the identification $x^+\sim x^+ + 2\pi$ is a null compactification throughout all of pure AdS$_7$, as is evident from the slicing (\ref{eq: AdS7 pp-wave slicing}).

In contrast, for the black hole solution (\ref{eq: BH plane wave metric}), $\partial_+$ is spacelike everywhere, with a norm-squared going like $\mathcal{O}(u^{-4})$ as we approach $u\to\infty$. Crucially, in the vicinity of the black hole horizon, the norm-squared can be made large in units of the Planck length as first noted in \cite{Maldacena:2008wh}, and thus we are able to trust semiclassical supergravity at large $N$.\\

Let us see in more detail why this is true. We compute everywhere in the black hole solution,
\begin{align}
  ||V||^2 = g_{++} &= \frac{\lambda^2}{g^2 }\frac{(1-g^2 y^2)^2(1-g^2 z^2)^2}{(1-a_1^2g^2)^2(1-a_2^2 g^2)^2} \frac{H^{-8/5}}{(u^2 + y^2)^2(u^2 + z^2)^2}\nn\\
  &\qquad \times\Big(\left(2m+q-(a_1-a_2)^2 g^2 q\right)(u^2 + y^2)(u^2 + z^2)-(a_1-a_2)^2 g^2 q^2\Big)
\end{align}
We do indeed see that $||V||^2$ vanishes like $u^{-4}$ as we approach the conformal boundary. Let us further focus on BPS black holes, for which $m,q$ can be replaced by functions of $a_1,a_2$, and let us consider the value of $||V||^2$ at the black hole horizon $u=u_0$. For brevity, let us further specialise to the case of equal angular momenta. Then, replacing $(y,z)$ with $(\theta_1,\theta_2)$ as in (\ref{eq: y,z angle replacement}) and setting
\begin{align}
  a_1 = a_2 =-\frac{1}{g}\left(\frac{1-\alpha^2}{1+\alpha^2}\right),\quad \alpha\in (0,1)
\end{align}
we find simply
\begin{align}
  \left.\frac{||V||^2}{l_p^2}\right|_{u=u_0} = 16\lambda^2(\pi N)^{2/3}\frac{\cos^2(\theta_2)(1-\alpha^2 \cos(2\theta_2))}{(1+\alpha^2)(1+\alpha^2 \sin(\theta_2)^2)^2}
  \label{eq: final length of circle}
\end{align}
where we have utilised (\ref{eq: AdS radius in Planck units}). Thus, for suitably large $N$ and $\lambda$ and away from\footnote{Note, there is a co-dimension $1$ hypersurface in the horizon, and indeed throughout the spacetime, at $\theta_2=\pi/2$ along which $||V||$ vanishes and thus the circle is null. This was also noted in \cite{Maldacena:2008wh}. Like those authors, we argue that this region of spacetime does not contribute appreciably to the thermodynamics, and thus can be neglected.} $\theta_2=\pi/2$, the circle is large in Planck units near the horizon. Below, we will use this to derive a classical gravity regime for $N$ and $K$.

\subsection{Thermodynamics}

Let us consider the black hole solution (\ref{eq: BH plane wave metric}) in the BPS limit, in which $\delta$ and and $m$ are fixed as functions of the remaining parameters $(\lambda, a_1, a_2)$. In additional to its Bekenstein-Hawking entropy $\ent_\text{BH}=A/4$ for horizon area $A$, we have thermodynamics quantities $\{K, \Hosc, J_1, J_2\}$ corresponding to a choice of commuting conformal Killing vectors of the boundary pp-wave metric, as well as a pair of equal R-charges $Q=Q_1=Q_2$.

The values of these quantities are
\begin{align}
  \ent_\text{BH}			&=	\frac{1}{G_N^{(7)}}\frac{\pi^3\lambda(a_1+a_2)(1-a_1g)(1-a_2g)(a_1 + a_2 - 2 a_1 a_2 g)}{2 g^4 (1+ a_1 g)(1+a_2 g)(2-a_1 g - a_2 g)^2 u_0}			&> 0	\nn\\
  K		&=	-\frac{1}{G_N^{(7)}}\frac{\pi^2 \lambda^2 (a_1 + a_2)(1-a_1 g)^2(1-a_2g)^2}{ g^4 (1+ a_1 g)(1+ a_2 g)(2-a_1 g-a_2 g)^2}		&> 0	\nn\\
  \Hosc			&= -\frac{1}{G_N^{(7)}}\frac{\pi^2 \lambda (a_1+a_2)(1-a_1 g)(1-a_2 g)(2-a_1^2 g^2-a_2^2 g^2)}{g^4(1+a_1 g)^2(1+a_2 g)^2 (2-a_1 g-a_2 g)^2}		&> 0			\nn\\
  J_1		&= 	 \frac{1}{G_N^{(7)}}\frac{\pi^2 \lambda (a_1 + a_2)(1-a_1 g)(1-a_2 g)(3a_1 + a_2 - a_1^2 g - 3a_1 a_2 g)}{4 g^3(1+ a_1 g)^2(1+a_2 g)(2- a_1 g - a_2 g)^2}			\nn\\
  J_2		&=  \frac{1}{G_N^{(7)}}\frac{\pi^2 \lambda (a_1 + a_2)(1-a_1 g)(1-a_2 g)(3a_2 + a_1 - a_2^2 g - 3a_1 a_2 g)}{4 g^3(1+ a_2 g)^2(1+a_1 g)(2- a_1 g - a_2 g)^2}						\nn\\
  Q			&=  -\frac{1}{G_N^{(7)}}\frac{\pi^2 \lambda (a_1 + a_2)(1-a_1 g)(1-a_2 g)}{4 g^4 (1+ a_1 g)(1+ a_2 g)(2- a_1 g - a_2 g)} &> 0	
\label{eq: final thermodynamics}
\end{align}
where the definite signs follow from (\ref{eq: a1 a2 bounds}). Additionally, while $J_1,J_2$ do not individually have definite sign over the physical parameter space, we do have $J_1+J_2>0$ for all physical solutions. $G_N^{(7)}$ denotes the seven-dimensional Newton's constant, related to the eleven-dimensional Newton's constant by $G_N^{(7)}=\text{Vol}(S^4)^{-1}G_N^{(11)}=6\pi^5l_p^5(\pi N)^{-4/3}$, where $G_N^{(11)}=16\pi^7 l_p^9$. The holographic relation (\ref{eq: AdS radius in Planck units}) then tells us that
\begin{align}
  \frac{R_\text{AdS}^5}{G_N^{(7)}} = \frac{1}{g^5 G_N^{(7)}} = \frac{16N^3}{3\pi^2}
\end{align}
 These thermodynamics can be deduced \cite{GravityFollowUp} from a Penrose limit of the known thermodynamics of the AdS$_7$ black hole solution of \cite{Chow:2007ts}. They satisfy the BPS condition
\begin{align}
  \Hosc-J_1 - J_2 - 4Q = 0
  \label{eq: final BH BPS condition}
\end{align}
precisely matching (\ref{bps1}) as expected. They additionally satisfy a non-linear constraint, whose origin can be traced back to the condition (\ref{eq: q constraint}) on parameters; this constraint is compactly expressed by the determination of $Q$ in terms of $J_1,J_2$ and $K$ as explained in the main text. Then, $\ent_\text{BH}$ can be expressed as a function of only $J_1,J_2$ and $K$, as appears in the main text.

\subsection{Range of classical validity}

As discussed in the main text, there are a pair of conditions under which M-theory is well described by supergravity. Let us check these now.
\paragraph{}
Firstly, we should have that the AdS radius and the radius of the internal $S^4$ are both large in Planck units. We have
\begin{align}
  \frac{R_\text{AdS}}{l_p} = 2 \frac{R_{S^4}}{l_p} = 2 (\pi N)^{1/3}
\end{align}
and thus we require $N>>1$. 
\paragraph{}
Secondly, we must ensure that the compact circle is large in Planck units, at least at the horizon. From our expression (\ref{eq: final length of circle}), we find that it is sufficient that the parameter $\lambda$ satisfies
\begin{align}
  \lambda N^{1/3}>> 1
  \label{eq: lambda condition}
\end{align}
Let us then convert this to statement about the charges $\{K,J_1,J_2,Q\}$. We can write
\begin{align}
  K 		&= \lambda^2 N^3 F_K(a_1 g,a_2 g),		\nn\\
  J_1 		&= \lambda N^3 F_{J_1}(a_1 g, a_2 g),	\nn\\
  J_2   	&= \lambda N^3 F_{J_2}(a_1 g, a_2 g),	\nn\\
  Q		   	&= \lambda N^3 F_{Q}(a_1 g, a_2 g),
\end{align}
for some functions $F_K,F_{J_1},F_{J_2},F_Q$ of the $\mathcal{O}(1)$ parameters $a_1 g, a_2 g$. By then considering $\lambda$ satisfying (\ref{eq: lambda condition}) while holding $a_1,a_2$ fixed\footnote{More generally, one could consider regions of charge space corresponding to also taking some limit of $a_1,a_2$, towards either $0$ or $-1/g$. However, one can show that any such limit corresponds either to even faster growth for $\{K,J_1,J_2,Q\}$, or else a horizon area that becomes small in Planck units.}, we find that the full set of sufficient conditions for the classical approximation to hold near the horizon is recast as
\begin{align}
  K>> N^{7/3} >> 1
\end{align}
with $J_1,J_2,Q$ growing like
\begin{align}
  J_1 \sim \sqrt{\delta},\qquad J_2\sim \sqrt{\delta},\qquad Q\sim \sqrt{\delta}
\end{align}
where recall $\delta=KN^3/3$.
\section{Entropy from quantum mechanics}\label{app: index asymptotics}
\subsection{Nekrasov partition function}\label{app: Nek}
\paragraph{}
The Nekrasov partition function for five-dimensional gauge theory defined on $\mathbb{R}^{4}\times S^{1}$ will play a central role in our analysis. Specifically we consider the five dimensional lift of $\mathcal{N}=2$ supersymmetric Yang-Mills theory with gauge group $U(N)$ with a single hypermultiplet of mass $m$ in the adjoint representation. 
The resulting four-dimensional theory has an effective coupling constant $\tau$. 
The theory is defined in a an $\Omega$ background on $\mathbb{R}^{4}$ with rotation parameters $\epsilon_{1}$ and $\epsilon_{2}$. The Coulomb branch of the $5d$ theory is parameterized 
by complex coordinates $z_{i}$, for $i=1,,\ldots N$, on the maximal torus of $U(N)$. We will denote the $\{z_{i}\}$ collectively as $Z$. 
As before, we exponentiate these parameters to define, 
\begin{eqnarray}
t:=\exp(-\epsilon_{+})\qquad{} &  \qquad{} x:= \exp(-\epsilon_{-}) \qquad{} & \qquad{} q_{\tau}:=\exp(2\pi i \tau) = \exp(-\beta) \nonumber \\
y:=\exp(-m) \qquad{} & \qquad{} z_{i}:=\exp(-\sum_{a=1}^{i-1} v_{a} )\qquad & \label{params}
\end{eqnarray}
where $\mathbf{v}=(v_{1},v_{2},\ldots,v_{N-1})$ is the $SU(N)$ chemical potential introduced in Section 2. As above we have $\epsilon_{\pm}:=(\epsilon_{1}\pm\epsilon_{2})/2$. 
The various partition function we discuss below will each be defined either as functions of the fugacities $\{t,x,y,q_{\tau};Z\}$ or as functions of the chemical potentials 
$\{\epsilon_{1},\epsilon_{2},m,\tau;\mathbf{v}\}$. The relations (\ref{params}) between the two sets of variables is assumed throughout in the following.    
\paragraph{}  
We define the Nekrasov partition function for this 5d gauge theory, 
\begin{eqnarray}
\mathcal{Z}^{U(N)}_{\rm Nek} & := & \mathcal{Z}^{U(N)}_{\rm Nek}\left[t,x,y,q_{\tau};Z\right]
\label{Nek1}
\end{eqnarray}
 as the product of perturbative and instanton contributions, 
\begin{eqnarray}
\mathcal{Z}^{U(N)}_{\rm Nek} & = & \mathcal{Z}^{U(N)}_{\rm pert}\, \mathcal{Z}^{U(N)}_{\rm inst}
\label{Nek2}
\end{eqnarray}
As we review below, only the instanton piece is directly related to the superconformal index we study. In fact the perturbative piece will also be relevant to our discussion once modular transformations are taken into account. 
Here we will use the slightly non-standard definition for the perturbative part introduced in \cite{KimNahmgoong}. This is given as,
\begin{eqnarray}
\mathcal{Z}^{U(N)}_{\rm pert} &  = & \hat{\mathcal{Z}}_{\rm pert}\,
\left[\mathcal{Z}^{U(1)}_{\rm pert}\right]^{N}
\nonumber 
\end{eqnarray}
with,
\begin{eqnarray} 
\hat{\mathcal{Z}}_{\rm pert} & := & {\rm Pexp}\left[\left(\sum_{i>j}\,\frac{Z_{i}}{Z_{j}}\right)\, \frac{\left[yt\right]\,\left[y/t\right]}{\left[tx\right]\,\left[t/x\right]} \right] \nonumber \\ 
\mathcal{Z}^{U(1)}_{\rm pert} & := & {\rm Pexp}\left[\frac{1}{2}\frac{\left[yx\right]\,\left[y/x\right]}{\left[tx\right]\,\left[t/x\right]}\right]
\label{zpert}
\end{eqnarray}
where, for any $X$ we define, 
\begin{eqnarray}
\left[ X\right] & := & \sqrt{X}-\frac{1}{\sqrt{X}} 
\nonumber 
\end{eqnarray}
Here the rational functions appearing in the argument of each Plethystic exponential are  expanded as an infinite Taylor series in the  variables $t$, $x$, $y$ and $z_{i}$. 
Different expansions are convergent in different chambers of the parameter space. 
Using the rules introduced above, we can 
then rewrite the Plethystic exponentials as infinite products. For example, for $t\leq 1$, we expand the argument in powers of $q_{1}=tx=\exp(-\epsilon_{1})$ and 
$q_{2}=t/x=\exp(-\epsilon_{2})$ to get, 
\begin{eqnarray} 
\hat{\mathcal{Z}}_{\rm pert}=\Pi^{(+)}\left[\epsilon_{1},\epsilon_{2}, m;\mathbf{v}\right]& := & \prod_{i>j=1}^{N}\, \prod_{l_{1},l_{2}=0}^{\infty}\,\,\frac{\left(1-\frac{z_{i}}{z_{j}}q_{1}^{l_{1}}q_{2}^{l_{2}}\right) 
\left(1-\frac{z_{i}}{z_{j}}t^{2}q_{1}^{l_{1}}q_{2}^{l_{2}}\right) }{\left(1-\frac{z_{i}}{z_{j}}ty^{-1}q_{1}^{l_{1}}q_{2}^{l_{2}}\right) 
\left(1-\frac{z_{i}}{z_{j}} ty q_{1}^{l_{1}}q_{2}^{l_{2}}\right) }
\label{infprod1}
\end{eqnarray}
On the other hand, for $|t|>1$, we can expand the argument of the Plethystic exponential in inverse powers of $q_{1}$ and $q_{t}$ to get, 
\begin{eqnarray} 
\hat{\mathcal{Z}}_{\rm pert}=\Pi^{(-)}\left[\epsilon_{1},\epsilon_{2}, m;\mathbf{v}\right] & := & \prod_{i>j=1}^{N}\, \prod_{l_{1},l_{2}=0}^{\infty}\,\,\frac{\left(1-\frac{z_{i}}{z_{j}}q_{1}^{-l_{1}}q_{2}^{-l_{2}}\right) 
\left(1-\frac{z_{i}}{z_{j}}t^{-2}q_{1}^{-l_{1}}q_{2}^{-l_{2}}\right) }{\left(1-\frac{z_{i}}{z_{j}}t^{-1}y q_{1}^{-l_{1}}q_{2}^{-l_{2}}\right) 
\left(1-\frac{z_{i}}{z_{j}} t^{-1}y^{-1} q_{1}^{-l_{1}}q_{2}^{-l_{2}}\right) } \nonumber 
\\
\label{infprod2}
\end{eqnarray}

The instanton part of the Nekrasov partition function is defined as an infinite sum over instanton number $K$,   
\begin{eqnarray}
\mathcal{Z}^{U(N)}_\text{inst} & = & \sum_{K=0}^{\infty} q_{\tau}^{K} \mathcal{Z}_{\rm inst}^{(K)} 
\label{inst1}
\end{eqnarray}
The rescaled coupling $\beta:=-2\pi i\tau$ serves as a chemical potential for instanton number $K$ and $q_{\tau}=\exp(-\beta)$ is the corresponding fugacity.
For each $K$ the contribution $\mathcal{Z}_{\rm inst}^{(K)}$ is expressed as a sum over $N$-coloured Young Tableaux,  
\begin{eqnarray}
\mathcal{Z}^{(K)}_{\rm inst}\left[t,x,y,;{Z}\right] & = & \sum_
{\vec{\lambda}\in \mathcal{P}^{N}: ||\vec{\lambda}||=K} \,\, \prod_{i,j=1}^{N}\,\, \prod_{s\in Y(\lambda^{(i)})}\,\,{\rm Pexp}\left(\frac{z_{i}}{z_{j}}\,t^{g_{ij}(s)}\,x^{f_{ij}(s)}\left[ty\right] \left[t/y\right]\right) \nonumber \\
\label{inst2}
\end{eqnarray}
with notation described in Appendix A.2. This yields a rational function of the fugacities which coincides precisely with the superconformal index of the instanton moduli space introduced in Section 2, 
\begin{eqnarray}
\mathcal{I}_{\rm SC}\left(\mathcal{M}_{K,N}\right) & = & \mathcal{Z}_{\rm inst}^{(K)} 
\label{indexvsnek}
\end{eqnarray}
We project onto the $\mathfrak{su}(n)$ singlet sector of the model by integrating against the Haar measure, 
\begin{eqnarray}
\int \, d\mu_{\rm Haar}[{Z}] & = & \left(\prod_{i=1}^{N} \,\frac{1}{2\pi i}\, \oint_{C}\, 
\frac{dz_{i}}{z_{i}}\right)\,\, \prod_{i\neq j}\, \left(1-\frac{z_{i}}{z_{j}}\right) 
\label{Haar}
\end{eqnarray}
Combining the above relations, the quantum mechanical index defined in the text can be expressed as, 
\begin{eqnarray}
\mathcal{I}_{\rm QM}\left[t,x,y,q_{\tau}\right] := 
\oint d\mu_{\rm Haar}\left[{Z}\right]\, \mathcal{Z}^{U(N)}_{\rm inst}\left[t,x,y,q_{\tau};{Z}\right] 
\label{indnek}
\end{eqnarray}
To ensure convergence of the $t$ expansion, at each instanton number we should work with $|t|<1$, keeping the other fugacities $x$ and $y$ on or near the unit circle. The Haar 
measure is also define above as an integration around the unit circle for each $z_{i}$.

\subsection{Elliptic resummation formula}
\paragraph{}
The explicit expression for $Z_{S}^{(\mathbf{n})}$ involves a sum over the set $S(\mathbf{n})$ of N-coloured partitions $\vec\lambda=(\lambda^{(1)},\ldots,\lambda^{(N)})$ with $|\lambda^{(i)}|=n_{a}$ for $i=a=1,2,\ldots,N-1$ and $|\lambda^{(N)}|=0$, 
\begin{eqnarray}
Z_{S}^{(\mathbf{n})} & = & \sum_{\vec{\lambda}\in S(\mathbf{n})}\,\prod_{i=1}^{N}\, \prod_{s\in Y(\lambda^{(i)})}\,\, \frac{t}{y}\,\,\frac{\vartheta\left(\tau | U_{1}^{(i)}(s)\right) \vartheta\left(\tau | U_{2}^{(i)}(s)\right)}{\vartheta\left(\tau | V_{1}^{(i)}(s)\right) \vartheta\left(\tau | V_{2}^{(i)}(s)\right)}
\label{elliptic}
\end{eqnarray}
For each box
$s\in Y(\lambda^{(i)})$, we define, 
\begin{eqnarray}
U_{1}^{(i)}(s):=\left(m-\epsilon_{+}+(l_{i+1}(s)+1)\epsilon_{2}-a_{i}(s)\epsilon_{1} 
\right)/2\pi i & &   V_{1}^{(i)}(s):=  \left(-l_{i}(s)\epsilon_{2}+(a_{i}(s)+1)\epsilon_{1} 
\right)/2\pi i   \nonumber \\ U_{2}^{(i)}(s):=
\left(m-\epsilon_{+}-l_{i-1}(s)\epsilon_{2}+(a_{i}(s)+1)\epsilon_{1} 
\right)/2\pi i
 &  &    V_{2}^{(i)}(s):=  \left((l_{i}(s)+1)\epsilon_{2}-a_{i}(s)\epsilon_{1} 
\right)/2\pi i  \nonumber 
\label{elliptic2}
\end{eqnarray}
and we extend the range of the index $i$ by the identification $\lambda^{(i\pm N)}:=\lambda^{(i)}$. 

\paragraph{}

The theta function $\vartheta(\tau|z)$ appearing in (\ref{elliptic}) is defined for all $\tau$, $z\in\mathbb{C}$ with ${\rm Im}\tau>0$ as, 
\begin{equation}
    \vartheta(\tau|z) = \sum_{n\in\mathbf{Z}} \exp((n^2 + n)i\pi \tau)(-1)^n\exp(2\pi i n z)
\end{equation}
It has a modularity property under $\tau\to -1/\tau, z\to z/\tau$:
\begin{equation}
    \vartheta(\tau|z) = \vartheta(-\frac{1}{\tau}|\frac{z}{\tau})\frac{1}{\sqrt{-i\tau}}\exp\left(\frac{(z + \tau/2 - 1/2)^2\pi}{i\tau}\right)
\end{equation}
The modular transformation property (\ref{key},\ref{kernel}) quoted in the text can be derived by applying this formula to each occurence of $\vartheta$ in (\ref{elliptic}). 

\subsection{Asymptotics of $Z_{S}$}

Our next aim is to find the asymptotic behaviour of $Z_{S}[\mathbf{v}]$ in the limit $\beta=-2\pi i \tau \rightarrow 0$. In particular the saddle point analysis will require us to take this limit with fixed complex values of the remaining chemical potentials. We will work indside the domain of analyticity of the index with ${\rm Re}[\beta]$, ${\rm Re}[\epsilon_{1}]$, ${\rm Re}[\epsilon_{2}]>0$. 
Note that the convegence of the sum (\ref{vevseries}) defining $Z_{S}$ typically requires 
additional conditions. However as explained in the text we assume that we can extend the definition of $Z_{S}$ to the whole domain by analytic continuation in $\beta$ so that the relation,
\begin{eqnarray}
Z_{S}\left[\epsilon_{1},\epsilon_{2},m,\tau;\mathbf{v}\right] & := & \frac{\mathcal{Z}^{U(N)}_{\rm Nek}}{\left[\mathcal{Z}^{U(1)}_{\rm Nek}\right]^{N}} \nonumber \\ 
& =& \hat{\mathcal{Z}}_{\rm pert}\,\frac{\mathcal{Z}^{U(N)}_{\rm inst}}{
\left[\mathcal{Z}^{U(1)}_{\rm inst}\right]^{N}} 
\label{equal}
\end{eqnarray}
is an equality between analytic functions. With this in mind, it is useful to extract the coeficients in the expansion (\ref{vevseries}) of the partition function and its dual by integrating the sum against the following measure, 
\begin{eqnarray}
\oint d\hat{\mu}\left[\mathbf{v}\right] & := & \prod_{a=1}^{N-1}\,\left( \frac{1}{2\pi i}\, 
\oint_{C_{a}}\, \frac{dw_{a}}{w_{a}}\right) 
\nonumber{}
\end{eqnarray} 
where, $w_{a}=\exp(-v_{a})$ for $a=1,2,\ldots,N-1$, and the contour $C_{a}$ is a small circle in the complex $w_{a}$-plane enclosing the origin.    
We then have, 
\begin{eqnarray}
Z^{(\mathbf{n})}_{S} & = & \oint d\hat{\mu}\left[\mathbf{v}\right] \exp\left(+\mathbf{n\cdot v}\right) Z_{S}\left[\mathbf{v}\right] 
\label{l1} 
\end{eqnarray}
and
\begin{eqnarray}
\tilde{Z}^{(\mathbf{n})}_{S} & = & \oint d\hat{\mu}\left[\mathbf{v}\right] \exp\left(+\mathbf{n\cdot v}\right) \tilde{Z}_{S}\left[\mathbf{v}\right] 
\label{l2}  
\end{eqnarray} 
with the S-duality relation $Z_{S}^{(\mathbf{n})} =  \mathcal{K}_{+}^{(\mathbf{n})}\,\tilde{Z}_{S}^{(\mathbf{n})}$.  
%
\paragraph{}
The first step is to find the asymptotics of 
the dual coefficients $\tilde{Z}_{S}^{(\mathbf{n)}}:=Z^{(\mathbf{n})}_{S}[\tilde{\epsilon}_{1},\tilde{\epsilon}_{2},\tilde{m},\tilde{\tau}]$. A direct approach to this problem starting from the explicit expression (\ref{elliptic},\ref{elliptic2}) is not straightforward for general values of the parameters so we will take a different approach.
We start from the equality (\ref{equal}) evaluated on the S-dualized parameters, 
\begin{eqnarray} 
\tilde{Z}_{S}[\mathbf{v}] & = & \tilde{\hat{\mathcal{Z}}}_{\rm pert}^{U(N)}\,\,\frac{
\tilde{\mathcal{Z}}^{U(N)}_{\rm inst}}{\left[
\tilde{\mathcal{Z}}^{U(1)}_{\rm inst}\right]^{N}} 
\nonumber 
\end{eqnarray}
where, as above, $\tilde{}$ denotes the replacement (\ref{rep}). Thus, for example, 
\begin{eqnarray}
\tilde{\mathcal{Z}}^{U(N)}_{\rm inst} & = & \mathcal{Z}^{U(N)}_{\rm inst}\left[\tilde{t},\tilde{x},\tilde{y},\tilde{q}_{\tau};{Z} \right] 
\nonumber 
\end{eqnarray}
with
\begin{eqnarray}  
\tilde{q}_{\tau}\,&:=& \,\exp\left(-\tilde{\beta}\right)\,=\,\exp\left(-\frac{4\pi^{2}}{\beta}\right) \nonumber \\ 
\tilde{y}\,&:=& \,\exp\left(-\tilde{m}\right)\,=\,\exp\left(-2\pi i \frac{m}{\beta}\right) \nonumber \\ 
\tilde{t}\, & := & \,\exp\left(-\tilde{\epsilon}_{+}\right) ,=\,\exp\left(-2\pi i \frac{\epsilon_{+}}{\beta}\right)\nonumber \\
\tilde{x} \,& := & \exp\left(-\tilde{\epsilon}_{-}\right),=\,\exp\left(-2\pi i \frac{\epsilon_{-}}{\beta}\right)
\end{eqnarray}
\paragraph{}
For small $\beta$, $|\tilde{q}_{\tau}|<<1$, and we have, 
\begin{eqnarray}
\tilde{Z}_{S} & = & \tilde{\hat{\mathcal{Z}}}^{U(N)}_{\rm pert}\,\,\left[\,1\,\,+\,\, O\left(\tilde{q}_{\tau}\right)\,\right] 
\nonumber{}
\end{eqnarray}
It is tempting to conclude that $\tilde{Z}_{S}$ can be replaced by $ \tilde{\hat{\mathcal{Z}}}_{\rm pert}$ which is in line with our expectation that the theory becomes weakly-coupled in the S-dual frame. In fact this conclusion depends sensitively on the values of the other parameters appearing in $\tilde{Z}_{S}$.
We will adopt this as a hypothesis and then describe some checks that this indeed the case for the saddle point values of the parameters. Making this assumption in (\ref{l1}) we have,  
\begin{eqnarray}
\tilde{Z}_{S}^{(\mathbf{n})} & \simeq & \int\, d\hat{\mu}\left[\mathbf{v}\right]\, 
\exp\left(+\,\mathbf{n}\cdot \mathbf{v}\right)\, \Pi^{(\pm)}\left[
\frac{\epsilon_{1}}{\tau}, \frac{\epsilon_{2}}{\tau},\frac{m}{\tau};\mathbf{v}\right] 
\label{intpm}
\end{eqnarray}
where $\Pi^{(\pm)}$ are the infinite products defined in (\ref{infprod1},\ref{infprod2}) above, choosing the 
"$+$" and "$-$" signs when ${\rm Re}[\epsilon_{+}/\tau]$ is positive and negative respectively. 
\paragraph{}
The integral (\ref{intpm}) can be evaluated formally as an infinite sum over the residues of the product $\Pi^{(\pm)}$. For each $\mathbf{n}\in \mathbb{Z}_{\geq 0}^{N-1}$, the leading term in the limit $\beta\rightarrow 0$, comes from the pole which maximises the value of  
\begin{eqnarray}
& {\rm Re}\left[\mathbf{n}\cdot \mathbf{v}\right] &
\end{eqnarray}
\paragraph{}
To investigate this further we must make some assumptions about the values of the other chemical potentials in the problem. Ultimately these values will be fixed by the saddle point equations studied in the next section. Although a more general analysis is possible, we will mainly focus on the special case of equal charges $Q_{1}=Q_{2}$ and equal angular momenta $J_{1}=J_{2}$ where the saddle point equations will impose $\epsilon_{1}=\epsilon_{2}=\epsilon_{+}$ and $m\simeq i \pi$. In the limit $\beta\rightarrow 0$ the dual parameters scale like, 
\begin{eqnarray}  
\tilde{q}_{\tau}\,:=\,\exp\left(-\tilde{\beta}\right)\,=\,\exp\left(-\frac{4\pi^{2}}{\beta}\right) & \rightarrow & 0 \nonumber \\ 
\tilde{y}\,:=\,\exp\left(-\tilde{m}\right)\,\simeq \,\exp\left(-\frac{2\pi^{2}}{\beta}\right) 
\,=\, \tilde{q}_{\tau}^{\frac{1}{2}} & \rightarrow & 0 \nonumber \\ 
\tilde{t}\,:=\,\exp\left(-\tilde{\epsilon}_{+}\right) & \rightarrow & 0\qquad{}\qquad{} \,\,{\rm Re}\left[\tilde{\epsilon}_{+}\right] > 0 \nonumber \\
& \rightarrow & \infty \qquad{}\qquad{} {\rm Re}\left[\tilde{\epsilon}_{+}\right] < 0  \label{limit2}
\end{eqnarray}
In this limit we find that the dominant pole lies at $v_{a}=\tilde{m}-\tilde{\epsilon}_{+}$ ($v_{a}=\tilde{m}+\tilde{\epsilon}_{+}$), with  
$a=1,2,\ldots,N-1$, for  ${\rm Re}\left[\tilde{\epsilon}_{+}\right]>0$  (${\rm Re}\left[\tilde{\epsilon}_{+}\right]<0$). Evaluating the residue of the integrand in (\ref{intpm}) at these points immediately gives 
\begin{eqnarray}
\log \tilde{Z}_{S}^{(\mathbf{n})} & \begin{array}{c} _{\beta\rightarrow 0} \\ \sim  \\ \,\,\end{array} & \frac{m \mp \epsilon_{+}}{\tau}\sum_{a=1}^{n}\,n_{a}
\label{intpm2}
\end{eqnarray}
with the $\mp$ sign depending on the sign of ${\rm Re}[\epsilon_{+}/\tau]$. 
\paragraph{}
We have performed several checks that the asymptotics derived above are not altered by instanton corrections in powers of $\tilde{q}_{\tau}$. In particular one can check that instanton corrections to the residue of the poles considered above are indeed suppressed at the first few orders. It is also possible to check that no competing poles arise at any finite order in the instanton expansion. It would be interesting to investigate whether this remains true in the full non-perturbative partition function using its Gopakumar-Vafa reformulation as in \cite{Bootstrap,Duan1,Duan2}. 
\paragraph{}
The S-duality relation (\ref{key}) then allows us to determine the asymptotics of coefficients $Z_{S}^{(\mathbf{n})}$ as, 
\begin{eqnarray}
\log Z_{S}^{(\mathbf{n})} & \begin{array}{c} _{\beta\rightarrow 0} \\ \sim  \\ \,\,\end{array} & \frac{\epsilon_{1}\epsilon_{2}}{2\beta}\sum_{a,b=1}^{N-1} n_{a} \Omega_{ab} n_{b}\,\,-\,\,\frac{1}{\beta}\Delta^{(\pm)}_{1}\Delta^{(\pm)}_{2}\sum_{a=1}^{N-1} \,n_{a} \nonumber \\ \label{as}
\end{eqnarray}
with, 
\begin{eqnarray}
\Delta^{(\pm)}_{1} := m\pm \epsilon_{+} & \qquad{} \Delta^{(\pm)}_{2}:= 2\pi i -m \pm \epsilon_{+} 
\nonumber{} 
\end{eqnarray} 
Inserting this result in (\ref{vevseries}) we find tha the modular partition function $Z_{S}$ asymptotes term by term to the series, 
\begin{eqnarray}
Z_{S}\left[\mathbf{v}\right] & \sim & \sum_{\mathbf{n}\in \mathbb{Z}_{\geq 0}^{N-1}}\,\, 
\exp\left[\frac{\epsilon_{1}\epsilon_{2}}{2\beta}\sum_{a,b=1}^{N-1} n_{a} \Omega_{ab} n_{b}\,\,-\,\,\frac{1}{\beta}\sum_{a=1}^{N-1}
\left(\Delta^{(\pm)}_{1}\Delta^{(\pm)}_{2}+\beta v_{a}\right)n_{a}\right]
\label{conv}
\end{eqnarray}
As the Cartan matrix $\Omega_{ab}$ has positive eigenvalues, the series is convergent for ${\rm Re}[\epsilon_{1}\epsilon_{2}/\beta]<0$ and gives an asymptotic formula for $Z_{S}[\mathbf{v}]$ as $\beta\rightarrow 0$ in this case. 
\paragraph{}
To obtain a more general result we can return to equation (\ref{l1}) which can be analytically continued to define $Z_{S}[\mathbf{v}]$ beyond the domain of convergence of the original series (\ref{vevseries}). Substituting for $Z_{S}^{(\mathbf{n})}$ with its asymptotic form (\ref{as}), this equation then takes an asymptotic form, 
\begin{eqnarray}
\oint \,d\hat{\mu}[\mathbf{v}] \,\,\exp\left(+\mathbf{n}\cdot\mathbf{v}\right)\,Z_{S}[\mathbf{v}] & \begin{array}{c} _{\beta\rightarrow 0} \\ \sim  \\ \,\,\end{array} &  \exp\left[\frac{\epsilon_{1}\epsilon_{2}}{2\beta}\sum_{a,b=1}^{N-1} n_{a} \Omega_{ab} n_{b}\,\,-\,\,\frac{
\Delta^{(\pm)}_{1}\Delta^{(\pm)}_{2}}{\beta} \sum_{a=1}^{N-1}n_{a}  \right] \nonumber \\ 
\label{intzs}
\end{eqnarray}
Its easy to check that this constraint is satisfied for all $\mathbf{n}$ by the following asymptotic form for $Z_{S}[\mathbf{v}]$, 
\begin{eqnarray}
\log\, Z_{S}\left[\mathbf{v}\right]  & \begin{array}{c} _{\beta\rightarrow 0} \\ \sim  \\ \,\,\end{array} & \,-\frac{1}{2\epsilon_{1}\epsilon_{2}\beta}\,\, \sum_{a,b}^{N-1}\, 
\left(\Delta^{(\pm)}_{1}\Delta_{2}^{(\pm)}+\beta v_{a}\right) \Omega^{-1}_{ab} \left(\Delta^{(\pm)}_{1}\Delta_{2}^{(\pm)}+\beta v_{b}\right) \nonumber \\
\label{zsfinal}
\end{eqnarray}
In particular, using this asymptotic formula on the LHS of (\ref{intzs}), we can evaluate the resulting contour integral by steepest descent to reprodice the expression in the RHS. More precisely, the saddle-point approximation is valid provided the action is large at the stationary point.    
We find that there are then two overlapping regimes in which he  
asymptotic is valid. The first is $N>>1$, we will call this the {\em large-$N$ regime}. The second case of interest is the {\em Cardy regime} where $|\epsilon_{1}|$, $|\epsilon_{2}|<<1$ for arbitrary $N$. 
We will consider the asymptotics of the index in both these cases in the following. 
\paragraph{}
We also note that the same  asymptotic can be obtained \cite{LeeNahmgoong} in the case ${\rm Re}[\epsilon_{1}\epsilon_{2}/\beta]<0$ directly from the convergent series (\ref{conv}). In either the Cardy or large-$N$ regime, the discrete summation over $\mathbf{n}\in\mathbb{Z}^{N-1}_{\geq 0}$ can be traded for an integration over a continuous variable which can also be evaluated in saddle-point approximation giving. The above analysis suggests that the same asymptotic form (\ref{zsfinal}) is also valid for ${\rm Re}[\epsilon_{1}\epsilon_{2}/\beta]>0$. 
\paragraph{}
One additional subtle point relates to the condition determining the two cases $\pm$ in (\ref{zsfinal}). In our derivation, which of the two cases arise depends on the sign of the dual parameter $\mathcal{A}:={\rm Re}[\tilde{\epsilon}_{+}]=2\pi {\rm Im}[\epsilon_{+}/\beta]$. Thus the relevant condition depends on both $\beta$ and $\epsilon_{+}$. When ${\rm Re}[\epsilon_{1}\epsilon_{2}/\beta]<0$ 
and the series converges, we can check that the condition $\mathcal{A}>0$ (or $\mathcal{A}<0$) can always be replaced by a simpler one involving only $\epsilon_{+}$. In particular, it is easy to show that, for  ${\rm Re}[\epsilon_{1}\epsilon_{2}/\beta]<0$, we  have, ${\rm Sign}(\mathcal{A})={\rm Sign}({\rm Im}[\epsilon_{+}])$. In extending our analysis as described above, it will suffice to discuss analytic continuation in $\beta$ with the other parameters held fixed. Then the $\pm$ cases of the asymptotic (\ref{zsfinal}) are determined by the condition ${\rm Im}[\epsilon_{+}]<0$ (+ case) and ${\rm Im}[\epsilon_{+}]>0$ (- case) throughout the domain of analyticity in $\beta$.    
\subsection{Asymptotics of the Index}
\paragraph{}
Finally we can consider the implications of he above analysis for the asymptotics of the supersymmetric index $\mathcal{I}_{QM}$ as $\beta \rightarrow 0$. The above analysis provides an effective resummation of the instanton series for our quantum mechanical index. Inverting the relation (\ref{equal}), we can express the index in terms of the modular partition function $Z_{S}$. Suppressing the arguments of the various partition functions we have, 
\begin{eqnarray}
\mathcal{I}_{\rm QM} & = & \int\,d\mu_{\rm Haar}\,\mathcal{Z}^{U(N)}_{\rm inst} 
\nonumber \\ 
& =& \left[Z^{U(1)}_{\rm inst}\right]^{N}\int\,d\mu_{\rm Haar}\,\,\frac{1}{\hat{\mathcal{Z}}^{U(N)}_{\rm pert}}\, Z_{S} \nonumber 
\label{IQM2}
\end{eqnarray}
In either the Cardy or large-$N$ regime, we can use asymptotic form (\ref{zsfinal}) for $Z_{S}$ to get, 
\begin{eqnarray}
\log \mathcal{I}_{\rm QM}  & \begin{array}{c}_{\beta\rightarrow 0} \\  \sim \\ \, \end{array}  &  -\frac{N(N^{2}-1)}{24} \frac{\left(\Delta^{(\pm)}\Delta^{(\pm)}\right)^{2}}{\epsilon_{1}\epsilon_{2}\beta}\,\,+\,\,N\log 
\mathcal{Z}^{U(1)}_{\rm inst}\,\,+\,\,\log \mathcal{R} \,\,\
\label{asympform2}
\end{eqnarray}
where $\mathcal{R}$ is a residual integral, 
\begin{eqnarray}
\mathcal{R} &: = & \int d\mu_{\rm Haar}[\mathbf{v}] 
\frac{1}{\hat{\mathcal{Z}}_{\rm pert}}\,\exp\left[\frac{\Delta^{(\pm)}\Delta^{(\pm)}}{\epsilon_{1}\epsilon_{2}}\mathbf{\rho}\cdot\mathbf{v}\right] \times \exp\left[ 
-\frac{\beta}{2\epsilon_{1}\epsilon_{2}}\,\, \sum_{a,b}^{N-1}\, 
v_{a} \Omega^{-1}_{ab} v_{b}\right] 
\nonumber 
\end{eqnarray}
where $\mathbf{\rho}=(\rho_{1},\ldots,\rho_{n})$ is the $SU(N)$ Weyl vector with components $\rho_{a}=a(N-a)/2$, for $a=1,\ldots,N-1$ and we have used the identity $\sum_{a,b=1}^{N} \Omega^{-1}_{ab}=N^{2}(N-1)/12$. Further, in the limit 
$\beta\rightarrow 0$, the quadratic term of $O(\beta)$ can be dropped, and $\mathcal{R}$ 
reduces to, 
\begin{eqnarray}
\mathcal{R}_{0}[\epsilon_{1}\epsilon_{2},m] & = &  
\int d\mu_{\rm Haar}[\mathbf{v}] \left(\Pi^{(+)}[\epsilon_{1},\epsilon_{2},m;\mathbf{v}]\right)^{-1}
\,\,\exp\left[\frac{\Delta^{(\pm)}\Delta^{(\pm)}}{\epsilon_{1}\epsilon_{2}}\mathbf{\rho}\cdot\mathbf{v}\right]
\nonumber 
\end{eqnarray}
where $\Pi^{(+)}$ is the convergent infinite product representing $\hat{\mathcal{Z}}_{\rm pert}$ in the region ${\rm Re}[\epsilon_{+}]>0$. 
This integral is finite and independent of the coupling $\beta=-2\pi i\tau$. The quantity  $\mathcal{R}_{0}$ can be cast in the form of a perturbative index with scaling $\log\mathcal{R}_{0}\sim N^{2}$ at large $N$. 
\paragraph{}
Finally we can determine the leading asymptotics of our index in the two regimes of interest. For large $N$, the first term in (\ref{asympform2}) scales like $N^{3}$, the second term involving the $U(1)$ instanton partition function provides a correction of order $N$. The residual integral has a finite limit as $\beta\rightarrow 0$ and does not contribute to the leading order growth of the exponent which goes like $\beta^{-1}$. Thus for $N>>1$ we have,   
\begin{eqnarray}
\log \mathcal{I}_{\rm QM}  & \begin{array}{c}_{\beta\rightarrow 0} \\  \sim \\ \, \end{array}  &  -\frac{N^{3}}{24} \frac{\left(\Delta^{(\pm)}\Delta^{(\pm)}\right)^{2}}{\epsilon_{1}\epsilon_{2}\beta}
\label{asympform3}
\end{eqnarray}
\paragraph{}
In the Cardy regime,  $|\epsilon_{1}|$, $|\epsilon_{2}|<<1$, for fixed $N$, we have $\Delta^{(\pm)}_{1}\simeq m$ and $\Delta^{(\pm)}_{2}\simeq 2\pi i-m$. We can also use the asymptotic form (\ref{u1result}) for $\mathcal{Z}^{U(1)}_{\rm inst}$. Combining this with the first term in (\ref{asympform2}) we find, 
\begin{eqnarray}
\log \mathcal{I}_{\rm QM}  & \begin{array}{c}_{\beta\rightarrow 0} \\  \sim \\ \, \end{array}  &  -\frac{N^{3}}{24} \frac{m^{2}(2\pi i-m)^{2}}{\epsilon_{1}\epsilon_{2}\beta}
\label{asympform4}
\end{eqnarray}
The two regimes overlap and we can summarise our results by concluding that the asymptotic form (\ref{asympform3}) is valid in both the Cardy and large-$N$ regimes.

\subsection{Saddle-point analysis} 

We start from the representation of the index coefficients as a contour integral, 
\begin{eqnarray}
\mathcal{C}(L_{t},L_{x},L_{y},K) & = & \frac{1}{(2 \pi i)^{4}}\, \oint_{C_{\beta}} \, \frac{dq_{\tau}}{q_{\tau}^{K+1}}\,\oint_{C_{+}} \, \frac{dt}{t^{L_{t}+1}}\,
\oint_{C_{-}} \, \frac{dx}{x^{L_{x}+1}}\,\oint_{C_{m}} \, \frac{dy}{y^{L_{y}+1}}\,
\,\, \mathcal{I}_{\rm QM}\left[t,x,y,q_{\tau}\right] \nonumber \\
\label{int1}
\end{eqnarray}
where the contours $C_{\beta}$, $C_{+}$, $C_{-}$, $C_{m}$ are circles, centered at the origin of the complex $q_{\tau}$, $t$, $x$ $y$ planes of radii 
$R_{\beta}$, $R_{+}$, $R_{-}$ and $R_{m}$ respectively. The specific expansion which yields the superconformal index corresponds to choosing, 
\begin{eqnarray}
0<R_{\beta}<R_{+}<<1 & \qquad & R_{-}=R_{y}=1 
\nonumber 
\end{eqnarray}
This corresponds to expanding the rational function arising at each instanton number in powers of $t$ with $|\mathcal{M}|<1$ for arbitrary monomials of the form $\mathcal{M}=tx^{a}y^{b}\prod z_{i}^{c_{i}}$.  
\paragraph{}  
In this section we will show that integral (\ref{int1}) can be evaluate by saddle point in the scaling limit. Replacing the integrand with its asymptotic form near $\beta=0$ (for the case ${\rm Im}[\epsilon_{+}]<0$),   
\begin{eqnarray}
\mathcal{C}(L_{t},L_{x},L_{y},K) & = & \frac{1}{(2 \pi i)^{4}}\, \int_{\mu_{\beta}}^{\mu_{\beta}+2\pi i} \, 
d\beta \, \int_{\mu_{+}}^{\mu_{+}+2\pi i} \, d\epsilon_{+} \, \int_{\mu_{-}}^{\mu_{-}+2\pi i} \, d\epsilon_{-}\, \int_{\mu_{m}}^{2\pi i +\mu_{m}} \, dm\,
\,\, \exp\left(\mathcal{S}\right) \nonumber \\
\label{int2}
\end{eqnarray}
with $\mu_{+}=\log R_{+}$ with similar definitions for the other integration variables. Here the exponent is given as,  
\begin{eqnarray}
\mathcal{S}+2\pi i Q_{2} & = & -\frac{N^{3}}{24}\,\frac{\Delta_{1}^{2}\Delta_{2}^{2}}{\epsilon_{1}\epsilon_{2}\beta}\,+\,\Delta_{1} Q_{1}\,+\, \Delta_{2} Q_{2}\,+\,\epsilon_{1} J_{1}\,+\,\epsilon_{2} J_{2}\,+\,\beta K \nonumber \end{eqnarray}
where, as above, 
\begin{eqnarray}
L_{t} & = & Q_{1}+Q_{2}+J_{1}+J_{2}  \nonumber \\ 
L_{x} & = & J_{1}-J_{2} \nonumber \\
L_{y} & = & Q_{1}-Q_{2} \nonumber 
\end{eqnarray}
and
\begin{eqnarray}
\Delta_{1}=\Delta_{1}^{(+)}=m+\epsilon_{+} & \qquad{} & \Delta_{2}= \Delta_{2}^{(+)}=2\pi i - m +\epsilon_{+} \nonumber 
\end{eqnarray} 
\paragraph{}
To identify the dominant contribtion to the integral, investigate the stationary points of $\mathcal{S}$, satisfying, 
\begin{eqnarray}
\frac{\partial \mathcal{S}}{\partial \epsilon_{1}}\,=\,\frac{\partial \mathcal{S}}{\partial \epsilon_{2}}\,=\, \frac{\partial \mathcal{S}}{\partial m}\,=\,\frac{\partial \mathcal{S}}{\partial \beta} & = & 0 \nonumber \end{eqnarray}
\paragraph{}
The chemical potentials can be eliminated using the stationary conditions to obtain a quartic equation for the saddle-point action. 
\begin{eqnarray}
\mathcal{S}^{2}\left(\mathcal{S}+2\pi i(Q_{2}-Q_{1})\right)^{2} & = & 
\frac{8\pi^{2}}{3} N^{3} K 
\left(\mathcal{S}+2\pi i(J_{1}+Q_{2})\right)\left(\mathcal{S}+2\pi i(J_{2}+Q_{2})\right)
\label{quartic}
\end{eqnarray} 
\paragraph{}
For the rest of this section we will mainly focus on the simplest case of equal charges $Q_{1}=Q_{2}=Q$ and angular momenta $J_{1}=J_{2}=J$ and determine degeneracies as a function of $L:=2(J+Q)$. The following analysis can easily be generalised to the case of generic charges. In this special case the saddle point action solves 
\begin{eqnarray}
\mathcal{S}^{4} & = & 
\frac{8\pi^{2}}{3}N^{3}K\,
\left(\mathcal{S}+\pi i L\right)^{2}
\label{condition2}
\end{eqnarray}
To determine the dominant saddle-point, we define rescaled variables $s$ and $\lambda$ by,
\begin{eqnarray}
\mathcal{S}=2\pi \,\sqrt{K}\,\sqrt{\frac{N^{3}}{3}}\,s & \qquad{} & L= \sqrt{K}\,\sqrt{\frac{N^{3}}{3}}\,\lambda
\nonumber
\end{eqnarray}
and split the rescaled action into its real and imaginary part as $s=:x+iy$, $x$, $y\in \mathbb{R}$. The complex equation (\ref{condition2}), then yields a pair of real equations for $x$ and $y$,  
\begin{eqnarray}
x^{2} & = & y^{2} + \frac{y+\lambda/2}{y} \label{eqnsxy1} \end{eqnarray}
\begin{eqnarray}
2y^{4}+ y^{2}-\frac{\lambda^{2}}{4} & = & 0
\label{eqnsxy2}
\end{eqnarray}
Solutions with $x>0$ correspond to saddle-points which contribute exponentially large terms to the integral.
We find two real roots of (\ref{eqnsxy2}) $y=y^{(\pm)}=\pm y_{\star}$ with, 
\begin{eqnarray}
y_{\star} & = & \frac{1}{2}\left(-1+\sqrt{1+2\lambda^{2}}\right)^{\frac{1}{2}}\,\,>\,\,0
\nonumber{}
\end{eqnarray}
which give, 
\begin{eqnarray}
x=x^{(\pm)} & :=& \left(y_{\star}^{2}+\frac{y_{\star}\pm \lambda/2}{y_{\star}}\right)^{\frac{1}{2}}\,\,>\,\,0 \nonumber 
\end{eqnarray}
\paragraph{}
For either root, the corresponding saddle point is is located at $m=i \pi$ and, 
\begin{eqnarray}
\epsilon_{1}=\epsilon_{2}=\epsilon_{+} & = & i \pi \frac{x+iy}{x+i(y+\lambda)} \nonumber 
 \end{eqnarray}
thus we find,  
\begin{eqnarray}
{\rm Re}\left[\epsilon_{+}\right] & = & \pi \frac{\lambda x}{x^{2}+(y+\lambda)^{2}} \nonumber \\ 
{\rm Im}\left[\epsilon_{+}\right] & = & \pi \frac{x^{2}+y(y+\lambda/2)}
{x^{2}+(y+\lambda)^{2}} \nonumber 
\end{eqnarray}
Thus ${\rm Re}[\epsilon_{+}]$ is positive for both saddle-points, and it is easy to check using (\ref{eqnsxy1},\ref{eqnsxy2}) that ${\rm Im}[\epsilon_{+}]$ is positive (negative) for $(x,y)=(x^{(+)},y^{(+)})$, ($(x,y)=(x^{(-)},y^{(-)})$). It follows that only the "$-$" saddle-point lies within the domain of validity of the "$+$" case of the asymptotic form (\ref{zsfinal}). From now on we restrict our discussion to this saddle.  
\paragraph{}
At the saddle point, the coupling $\beta$ takes the value
\begin{eqnarray}
\beta & = & \pi \sqrt{\frac{N^{3}}{3K} }\,\frac{(x+iy)\left(x+i\left(y+i\lambda/2\right)\right) }{x+i(y+\lambda)} \nonumber \\ 
\end{eqnarray}
and therefore has real part, 
\begin{eqnarray}
{\rm Re}[\beta] & = & \pi \sqrt{\frac{N^{3}}{3K} }\frac{x(y+\lambda/2)}
{x^{2}+(y+\lambda)^{2}}\left(\frac{1}{y}+\lambda + 2 y\right)
\label{rpart} 
\end{eqnarray}
which remains positive provided that $\lambda>\lambda_{cr}:=2\sqrt{1+\sqrt{5}/2}\simeq 2.91$. Here the critical value, $\lambda_{cr}$, is determined by the largest zero of last factor in (\ref{rpart}). 
\paragraph{}
Combining the above results for $\beta$ and $\epsilon$, we therefore have $|t|<1$ and $|q_{\tau}|<1$ at the saddle point provided that\footnote{Our analysis cannot easily be extended to lower values of $\lambda$ without additional assumptions.}  
$\lambda>\lambda_{cr}$. Hence, in this range, the saddle-point lies inside the assumed domain of analyticity for the index.   
Note that one also has, 
\begin{eqnarray}
{\rm Re}\left[\frac{\epsilon^{2}}{\beta}\right] &  = &   -\pi\,\sqrt{\frac{3K}{N^{3}}}\, 
\frac{x(x^{2}+y^{2}-\lambda^{2}/2)}{(x^{2}+(y+\lambda)^{2})(x^{2}+(y+\lambda /2)^{2})}
\nonumber 
\end{eqnarray}
and one may check that this is always positive at the remaining saddle point. Thus the stationary point lies outside the region of convergence of the series (\ref{conv}). However, as discussed above the "$+$" case of the asymptotic (\ref{zsfinal}) still holds in this regime provided ${\rm Im}[\epsilon]<0$.   
\paragraph{}
Finally the analysis can be repeated with $\Delta^{(+)}_{i}$ replaced by $\Delta^{(-)}_{i}$, for $i=1,2$, and yields a second consistent saddle-point, with ${\rm Im}[\epsilon_{+}]>0$, related to the one discussed above by complex conjugation: $x\rightarrow x$, $y\rightarrow -y$ with $\lambda$ fixed. 
The two consistent saddle points each contribute an exponentially growing term to the integral. The resulting contribution to the asymptotics of the index coefficients can be written as, 
 \begin{eqnarray}
\log\, \left| \mathcal{C}[L_,0,0, K]\right| & \sim & 2\pi \sqrt{K}\,\sqrt{\frac{N^{3}}{12}}\,\left[ \sqrt{1+\sqrt{1+6L^{2}/KN^{3}}}-\sqrt{2}\right]
\label{logC2}
\end{eqnarray}
Where we take the $K\rightarrow \infty$ limit with $L\sim \sqrt{K}$. The above analysis is easily generalised to the case where $J_{1}\neq J_{2}$, $Q_{1}\neq Q_{2}$. The result has the scaling form, 
\begin{eqnarray}
\log\, \left| \mathcal{C}[L_{t},L_{x},0, K]\right| & \sim & 2\pi \sqrt{\delta}\,
\mathcal{F}_{\rm QM}\left[L_{t}/\sqrt{\delta},L_{x}/\sqrt{\delta},L_{y}/\sqrt{\delta}\right] 
\label{growth2}
\end{eqnarray}
with $\delta=KN^{3}/3$. The function $\mathcal{F}_{\rm QM}[u,v,w]$ is determined by the smaller positive root $s=s_{\star}$ of the quartic equation, 
\begin{eqnarray}
s^{2}(s-iw)^{2} & = & 2\left(s+i(u+v)/2\right)\left(s+i(u-v)/2\right)
\label{finalapp}
\end{eqnarray}
as $\mathcal{F}_{\rm QM}={\rm Re}\,s_{\star}$. 


Returning to the special case of equal charges and angular momenta, a more precise analysis is possible (for any $N$) in the Cardy regime corresponding to 
$L^{2}>>\delta/2=N^{3}K/6$. The complex conjugate pair of saddle-point actions can then be written as $\mathcal{S}(L) \pm i  
\mathcal{S}(L)$ with, 
\begin{eqnarray}
\mathcal{S}(L) & = & 2\pi\,\left(\frac{N^{3}}{24}L^{2}K\right)^{\frac{1}{4}}
\label{slintro}
\end{eqnarray}
Adding together the contribution of both saddle points, and including the prefactor from small fluctuations around the stationary point, 
we obtain a formula for the asymptotic growth of the index coefficients in the Cardy regime, 
\begin{eqnarray}
\mathcal{C}\left(L_,0,0,K\right) & \sim & 
\frac{3\sqrt{2}}{16\pi}\,\frac{\mathcal{S}(L)}{L^{2}K}\,\exp\left[\mathcal{S}(L)\right] \cos\left(  
\mathcal{S}(L)+\frac{\pi}{4} \right)
\nonumber 
\end{eqnarray}
with $\mathcal{S}(L)$ given by (\ref{sl}). In the Cardy regime, this formula 
holds for all $N$. In the case $N=1$, we have tested this formula against a direct expansion of the localisation formula on a computer up to $K\sim L\sim 60$, finding good agreement.

\bibliographystyle{JHEP}
\bibliography{BHfinal}

\end{document}